\documentclass[pdflatex,sn-mathphys-num]{sn-jnl}

\usepackage{graphicx}%
\usepackage{multirow}%
\usepackage{amsmath,amssymb,amsfonts}%
\usepackage{amsthm}%
\usepackage{mathrsfs}%
\usepackage[title]{appendix}%
\usepackage{xcolor}%
\usepackage{textcomp}%
\usepackage{manyfoot}%
\usepackage{booktabs}%
\usepackage{algorithm}%
\usepackage{algorithmicx}%
\usepackage{algpseudocode}%
\usepackage{listings}%
\usepackage{geometry}%
\usepackage[version=4]{mhchem}%
\usepackage{lineno}%
\usepackage{nomencl}
\usepackage{indentfirst}%
\usepackage{enumitem}%
\usepackage{makecell}%
\usepackage{pdflscape}
\usepackage{indentfirst}
\UseRawInputEncoding

\usepackage{subfiles}

\usepackage[globalcitecopy]{bibunits}
\usepackage{xr-hyper}
\usepackage{etoolbox} 
\makeatletter
\AtBeginEnvironment{bibunit}{%
  \edef\@extra@b@citeb{.\@bibunitname}%
  \edef\@extra@binfo{.\@bibunitname}%
}
\makeatother

\makeatletter
\def\addressfont{\reset@font\fontsize{9bp}{10bp}\selectfont\titraggedcenter}%

\makeatother

\begin{document}

\begin{bibunit}[sn-mathphys-num]


\title[Article Title]{Satellite-based emissions estimate indicates progress toward China's methane mitigation goals}
\author*[1,2]{\fnm{Ziting}   \sur{Huang}} \email{zhuang51@jhu.edu}
\author[1]{\fnm{Ao}         \sur{Chen}}
\author[1]{\fnm{Leyang}      \sur{Feng}}
\author[1]{\fnm{William S.}  \sur{Daniels}}
\author[1,3,4]{\fnm{Dylan C.}    \sur{Gaeta}}
\author[1]{\fnm{Kristan L.}  \sur{Morgan}}
\author[5]{\fnm{Lee T.}      \sur{Murray}}
\author[6]{\fnm{Xueying}      \sur{Yu}}
\author[1,2]{\fnm{Benjamin F.} \sur{Hobbs}}
\author*[1]{\fnm{Scot M.} \sur{Miller}} \email{smill191@jhu.edu}
\affil[1]{\orgdiv{Department of Environmental Health and Engineering}, \orgname{Johns Hopkins University}, \orgaddress{\street{3400 N Charles St}, \city{Baltimore}, \postcode{21218}, \state{MD}, \country{USA}}}
\affil[2]{\orgdiv{Ralph O'Connor Sustainable Energy Institute}, \orgname{Johns Hopkins University}, \orgaddress{3400 N Charles St}, \city{Baltimore}, \postcode{21218}, \state{MD}, \country{USA}}
\affil[3]{\orgdiv{Cooperative Institute for Research in Environmental Sciences}, \orgname{University of Colorado Boulder}, \orgaddress{\street{1665 Central Campus Mall 216 UCB}, \city{Boulder}, \postcode{80309}, \state{CO}, \country{USA}}}
\affil[4]{\orgdiv{Global Monitoring Laboratory}, \orgname{National Oceanic and Atmospheric Administration}, \orgaddress{\street{325 Broadway}, \city{Boulder}, \postcode{80305}, \state{CO}, \country{USA}}}
\affil[5]{\orgdiv{Department of Earth and Environmental Sciences}, \orgname{University of Rochester}, \orgaddress{\street{120 Trustee Rd}, \city{Rochester}, \postcode{14620}, \state{NY}, \country{USA}}}
\affil[6]{\orgdiv{Atmospheric Sciences Research Center}, \orgname{The State University of New York at Albany}, \orgaddress{\street{1400 Washington Avenue}, \city{Albany}, \postcode{12222}, \state{NY}, \country{USA}}}

\date{Jan 2026}

\abstract{
China emits the most methane of any country worldwide, but there are large uncertainties in recent emissions trends, sources, and the potential impacts of policy actions. This study focuses on a period when the government initiated ambitious methane control efforts, linking sectoral policies with atmospheric evidence on sectoral, sub-national, and seasonal emissions during 2019--2024. We quantify daily methane emissions from China using a regional atmospheric inverse model with TROPOMI satellite observations. Our results reveal an average methane emissions increase rate of 0.3 Tg yr$^{-2}$ in Eastern \& Central China, likely a milder trend than in the 2010s. Coal industry methane emissions intensity declined for the first time (-3.2\% yr$^{-1}$) despite rising production, possibly associated with diverse policy instruments, mandates, and incentives. We further highlight two emerging challenges for future mitigation: leaks from expanding urban gas use amid the energy transition and rising agricultural emission yet with substantial uncertainty in estimates. Lastly, declining emissions intensity of coal mines points to the future role of targeted mandates and incentives in encouraging methane reduction for other sectors.
}
\keywords{methane emissions, satellite inverse modeling, China, mitigation policy}



\maketitle

\newpage


\section{Introduction}\label{sec1}

China has notably intensified methane (\ce{CH_4}) mitigation efforts since COP26 \cite{IEANationalActionPlan_2024, GlasgowDeclaration2021}. It is the largest anthropogenic methane emitter among all countries for decades, sharing 10\%--16\% of annual global methane emissions \cite{crippa_insights_2024,CAMS_v23r1}. Since 2010, energy has likely overtaken agriculture as the dominant anthropogenic methane source in China, followed by the waste sector, with a respective share of 42\%, 34\%, and 19\% during 2010--2024 \cite{crippa_insights_2024}. Over the past decade, China's sectoral methane control policies also evolved with broader environmental and climate goals, moving beyond the traditional focus on safety and resource conservation \cite{zhu2025comparative}. In 2023, the government issued the first national policy agenda on systematic methane mitigation, \textit{Methane Emissions Control Action Plan (``Action Plan")}, with quantified methane management targets in three areas by 2025: (a) coal mine methane (CMM) in energy, (b) animal manure in agriculture, and (c) urban waste, including household waste and sewage sludge\cite{zhu2025comparative, methaneactionplan_2023}. This mitigation ambition requires extensive monitoring, rigorous accounting, and modeling to better characterize local methane sources. \\

This mitigation ambition requires extensive monitoring and rigorous accounting to better characterize local methane sources, yet China currently has a very limited network of in-situ, atmospheric greenhouse gas sensors. As of 2025, there are fewer than ten methane surface observation sites within the country \cite{ChinaGHGBulletin2025, GAWsearch, zhang2023monitoringsites}. Waliguan (WLG), the site with the longest operational record in China, generates $\sim$100 surface methane measurements per year on the Tibetan plateau \cite{NOAA_WLG, liu2021measurement}. By contrast, greenhouse gas satellites provide thousands to millions of column-averaged methane (\ce{XCH_4}) observations across China each year, such as Greenhouse gases Observing SATellite (GOSAT) and TROPOspheric Monitoring Instrument (TROPOMI) \cite{ohyama2024gosat, TROPOMI_manual2024}. Due to the wide spatial coverage, satellite retrievals have become a useful data product to help infer methane emissions for global regions since 2010, especially for areas with poor surface measurements. \\


Inverse modeling studies incorporate atmospheric measurements as complementary constraints to better quantify the scale and distribution of emissions, potentially improving inventories when inverse estimates disagree. Comparatively, emission inventories derive emissions estimates using assumptions on activities and emission factors. For example, Emissions Database for Global Atmospheric Research (EDGAR) regularly update methodology and activity data to reconstruct a high-resolution spatial profile (0.1$^{\circ}\times$0.1$^{\circ}$) for methane emissions with seasonal variations \cite{crippa_insights_2024}. However, inverse modeling studies present $>$10\% (7--12 Tg yr$^{-1}$) lower estimates than EDGAR v6.0 for total anthropogenic methane emissions in China during the 2010s (SI Figure \ref{fig:SI-uncertainty2010-2019}) \cite{miller_chinas_2019, zhang_observed_2022, zhao_slowdown_2024, minx2021comprehensive}. In turn, understanding this discrepancy can support a closed-loop feedback process to refine inventory assumptions. The latest EDGAR 2025 version reduces China's emissions by an average of 15.9 Tg yr$^{-1}$, compared to the earlier EDGAR v6.0, for overlapping years since 2010, mainly due to improved methods to calculate emissions from fossil fuels ($-$5 Tg yr$^{-1}$), rice cultivation ($-$5.9 Tg yr$^{-1}$), and wastewater treatment/discharge ($-$4.3 Tg yr$^{-1}$) \cite{Crippa2025GHG,Crippa2026_EDGARv60, minx2021comprehensive, crippa_insights_2024}. For inventory quality assurance, the Intergovernmental Panel on Climate Change (IPCC) also provides guidelines for member countries to benchmark inventories against national/sub-national emissions and trends derived from inverse models \cite{IPCCGuidlines2019Refine-Overview, IPCCGuidlines2019Refine-v1-cha6}. \\

Validating emissions trends is one particular area where inverse modeling studies can help inform future inventories for China. Current studies indicate a steady, rapid increase in methane emissions from China in the 2000s, from 1.0 to 1.2 Tg yr$^{-2}$ (e.g., \cite{thompson_methane_2015, bergamaschi_high-resolution_2022, bruhwiler_carbontracker-ch4_2014}). For the subsequent decade (2010s), satellite-based studies report a similar but slightly wider range of trend estimates, 0.4--1.1 Tg yr$^{-2}$ (SI Figure \ref{fig:SI-uncertainty2010-2019}) \cite{miller_chinas_2019, zhang_observed_2022, zhao_slowdown_2024}. These trend estimates broadly match emission inventories but highlight different growth periods in China: EDGAR 2025 version shows a surge ($\sim$0.8 Tg y$^{-2}$) from 2000 to the early 2010s, followed by a slight increase (0.1 Tg y$^{-2}$) in the late 2010s, with an overall 2010s trend of 0.2 Tg y$^{-2}$ \cite{crippa_insights_2024}. By comparison, the earlier EDGAR v6.0 exhibits a sharper rising trend (1.1--1.4 Tg y$^{-2}$) in 2000--2015 then a decline (-0.3 Tg y$^{-2}$) in 2015--2018 \cite{Crippa2026_EDGARv60, crippa_insights_2024}. Additionally, different inverse modeling studies attribute the 2010s trend to differing sources. Existing satellite-based studies agree that energy production contributed to the observed increase in methane emissions, particularly fugitive emissions from coal mining activities and abandoned coal mines (e.g., \cite{miller_chinas_2019, sheng_sustained_2021}). These studies, however, diverge on whether other sectors also played a role in the additional increase, including natural gas use, rice cultivation, and aquaculture farming \cite{miller_chinas_2019, sheng_sustained_2021, zhao_slowdown_2024, wang2022atmospheric_ng_use}. \\ 

It is unclear how methane emissions from China have been changing during the early 2020s, despite numerous inverse trend estimates for the 2000s and 2010s. Yet, it is pivotal to quantify these recent trends, given the evolving methane policy landscape in China. Here, we estimate anthropogenic methane emissions across China for years 2019--2024 using inverse modeling and TROPOspheric Monitoring Instrument (TROPOMI) atmospheric measurements. TROPOMI is a satellite instrument with the densest observations due to daily, global coverage (e.g., \cite{TROPOMI_manual2024}). 
We characterize methane emissions by sector (Section \ref{subsec2}), sub-nation (Section \ref{subsec3}), and season (Section \ref{subsec4}). We further combine production activities to estimate emissions intensity over time, and explore possible links with concurrent mitigation policies in China. In addition, we compare emissions estimates against previous global and regional studies to investigate the drivers of these differences. In this study, we mainly discuss the findings in Eastern \& Central China (east of the Tibetan Plateau), due to suspected retrieval biases and low signal-to-noise ratio of TROPOMI observations in Western China (with details in Section \ref{sec4}). Our study region is representative, accounting for $\sim$93\% of China's emissions, based on inventories (see divisions in SI Figure \ref{fig:regioncat}). \\

\section{Results and Discussion}\label{sec2}

\subsection{China tempered its methane emission growth}\label{subsec1}

We estimate average methane emissions in Eastern \& Central China during 2019--2024 to be 53.3 Tg yr$^{-1}$ (47.4 Tg yr$^{-1}$ from anthropogenic sources), with ensemble posterior estimates of 50.7--55.9 Tg yr$^{-1}$ (45.6--49.8 Tg yr$^{-1}$ anthropogenic). Annual ensemble mean estimates are on average lower than emission inventories by 7\% overall and 13\% for anthropogenic emissions. 
For comparison, our estimated emissions fall between estimates from global studies by Copernicus Atmosphere Monitoring Service (CAMS) v24r1 (58.3--68.1 Tg yr$^{-1}$) \cite{CAMS_description_2025} and by Pendergrass et al. (47--48.3 Tg yr$^{-1}$) \cite{pendergrass2025trends}, based on corresponding output files. 
Compared to global inverse models at coarser resolution, our finer-resolution regional inverse model can capture more spatial and temporal variability in local emissions, likely reducing aggregation biases from meteorological fields and improving source attribution. Our estimates are also close to the mean estimate (50.9 Tg yr$^{-1}$) from studies with Greenhouse gases Observing SATellite (GOSAT) observations for the 2010s (Figure \ref{fig1}A) \cite{zhang_observed_2022, CAMS_evaluation_2022}. \\ 

Total methane emissions grow at a milder rate (0.3 Tg yr$^{-2}$) in our study region, compared to the 2010s (Figure \ref{fig1}B). Our anthropogenic emissions exhibit the same consistent, increasing trend. Our linear trend in total methane emissions matches CAMS v24r1 (0.2 Tg yr$^{-2}$) but differs from the negative trend ($-$0.3 Tg yr$^{-2}$) of Pendergrass et al. for the same region. Overall, all post-2019 growth estimates of Eastern \& Central China, including ours, are lower than the national trends reported by regional GOSAT-based studies for the early 2010s \cite{zhang_observed_2022, miller_chinas_2019, zhao_slowdown_2024}. Our regional trends are comparable to nationwide trends from previous studies, as provinces in our study region accounted for $\sim$92\% of the national emissions trend in 2010--2017, based on the sub-national analysis by Zhang et al. \cite{zhang_observed_2022}. Additionally, our finding about slower growth also aligns with estimates in a previous study by Zhao et al. (0.1$\pm$0.3 Tg yr$^{-2}$) for the late 2010s (2016--2021) in China \cite{zhao_slowdown_2024}. \\ 

The moderate anthropogenic methane trend likely reflects slowing coal methane emissions and continued growth from livestock emissions (middle row of Figure \ref{fig2}). During 2019--2024, energy sources show mild increases in methane emissions, with an overall linear annual growth rate of 0.12 Tg yr$^{-2}$ (1\% yr$^{-1}$) comprised of 0.09 Tg yr$^{-2}$ from coal and 0.03 Tg yr$^{-2}$ from oil and gas (O \& G). Our inferred trend in post-2019 fossil fuel emissions is at the lower end of the estimates by earlier regional inverse modeling studies for the 2010s (0.12--0.25 Tg yr$^{-2}$) \cite{miller_chinas_2019, zhang_observed_2022, zhao_slowdown_2024}. Urban-related sources do not exhibit noticeable increases (0.01 Tg yr$^{-2}$, 0.2\% yr$^{-1}$). For agricultural emissions, we find a steady, increasing trend in livestock farming (0.34 Tg yr$^{-2}$, 3.2\% yr$^{-1}$), while a fluctuating, decreasing trend in rice cultivation (-0.19 Tg yr$^{-2}$, $-$2.1\% yr$^{-1}$). 
In general, sectoral emission trends in China also mirror global developments: globally, coal-related emissions remain relatively flat, while livestock emissions continue to rise (+2.8\% yr$^{-1}$), based on a global inverse modeling analysis by He et al. \cite{he2025attributing}.

\subsection{Coal sector exhibits a declining emission intensity} \label{subsec2}

Among all anthropogenic sources, the coal industry shows the sharpest negative trend in emissions intensity (bottom row of Figure \ref{fig2}). Coal industry emissions intensity declines linearly at -3.2\% yr$^{-1}$, in contrast to an earlier flat trend estimated for the 2010s by Zhang et al. \cite{zhang_observed_2022}. Despite the sustained domestic raw coal production growth (4.1\% yr$^{-1}$), we find that coal mine methane (CMM) emissions remained relatively stable in the same period (0.8\% yr$^{-1}$). CMM, formed during coalification and trapped within coal seams and surrounding rock strata, is released during and after mining activities, due to pressure changes and ventilation for safety (e.g., \cite{EPA_CMM, IEA_CMM}). Generally, the volume of CMM emitted from mines depends on mine type (underground or surface), coal type (lignite, subbituminous, bituminous, or anthracite), coal production rate, as well as site-specific requirements on mine gas drainage, ventilation, and re-utilization (e.g., \cite{GMI_BasicsCMM}). In China, the reduced coal methane emissions intensity in 2019--2024 is likely the result of two factors:

\begin{enumerate}[label=(\alph*)]
    \item \textbf{Mining shift toward regions with low-concentration mine gas.} Our results align with a recent bottom-up study using bootstrap simulations by Zhang et al. (2025), which suggests that China has a lower production-weighted average emission factor in 2023 than in 2011, due to the regional shift of coal production \cite{zhang2025regionalshift}. In China, the coal production share from lower-emitting surface mines increased from 17\% in 2019 to 25\% in 2023 \cite{ChinaEnergyData-2023, ChinaCoalYB-2023}. Northern and northwestern provinces, with more surface and low-emitting mines, further scaled up raw coal production in 2024 compared to 2019, including Shanxi (+31\%), Inner Mongolia (+25\%), Shaanxi (+23\%), and Xinjiang (+127\%), partly related to the new north-south coal railway (operating since 2019) \cite{NBS-coal-monthly,NBS-coalmine}. 
    Comparatively, many high-emitting, small underground mines closed in the southwestern mountain regions, while others shrunk production, especially in Sichuan (-34\% drop in coal production from 2019) \cite{NBS-coal-monthly,NBS-coalmine, NRDC_NationalMineClosure2019, NRDC_SichuanMineClosure2016}. 
    \item  \textbf{A combination of mandates and incentives on CMM control.} As of 2024, China continued regulating high-concentration coal mines ($\geq 30\%$ methane content) in three ways: (i) license qualified mines before production, (ii) close unsafe, inefficient mines that violate emission standards, and (iii) penalize operators for illegal acts and accidents \cite{NEA2005_coalmine_methane}. Additionally, mine owners who utilize CMM receive two voluntary emission-reduction incentives: exemption from prospecting and mining royalties through 2020 \cite{NEA2005_coalmine_methane}, and utilization subsidies at the central and provincial levels \cite{NEA2007_cbm_subsidy, MOF2016_cbm_subsidy_adjustment}. Since 2019, the central government has adopted a performance-based subsidy system (\textyen 0.3--0.6$\mathrm{/m^3}$), replacing the previous uniform, fixed per-unit production subsidy (\textyen $0.3 \mathrm{/m^3}$), to encourage producing unconventional gas, including coal-bed methane (CBM) \cite{2019specialfund_unconventionalgas}. The new scheme increases incentives for coal mine operators to capture and reuse CMM by flexibly allocating annual subsidies (US\$300--600M), based on individual contribution to incremental gas output 
    \cite{2021NEA_response, 2021subsidy_issue, 2019-2024subsidy_earlyissue}. The government further differentiates subsidies with reward factor according to year-on-year production growth of gas producers, assigning higher factor to CBM/CMM or gas used for winter heating \cite{2019specialfund_unconventionalgas, ChinaEnergyNews2025_CBMsubsidy}. This reformed subsidy policy boosted nationwide CMM utilization from 4.7 billion $\mathrm{m^3}$ in 2018 to 5.7 billion $\mathrm{m^3}$ in 2020, raising utilization rate to 45\% \cite{CMMutilizationSX_unece_2024, MEE_Utilization2020_2023}. Overall, dynamic subsidies likely motivate more CMM production from mines with lower mitigation costs, moving China closer to the 6 billion $\mathrm{m^3}$ CMM utilization target by 2025, equivalent to avoiding 37\% of our estimated CMM emissions. 

\end{enumerate}

Other sectors also likely experience reduced methane emissions intensity albeit with greater uncertainty, except for the livestock sector. We further explore the plausibility of these trends by cross-checking activity data, industrial/commercial reports, and research literature. In the O\&G sector, emissions intensity falls by 1.7\% yr$^{-1}$ despite rising O\&G production, possibly due to direct methane-control policies, including mandatory reporting and monitoring of field emissions since 2019 \cite{OGIndustry_2019}. This decline aligns with a 44\% drop in intensity between 2019 and 2024 ($\sim$7.6\% y$^{-1}$) reported by the China National Petroleum Corporation (CNPC), China's largest O\&G producer \cite{CNPCreport2024}. For urban-related sources, we use waste indicators as proxies and find a possible decline in emissions intensity from waste and wastewater management activities (-2.7\% yr$^{-1}$). We also observe a potentially slight reduction in the rice cultivation emissions intensity (-1.7\% yr$^{-1}$). These intensity trends coincide with indirect, conservation-focused sectoral policies: (a) zero-waste city initiatives \cite{ZeroWasteCity2021} and subsidized waste-to-energy (WtE) power \cite{NRDC_subsidyWtE_2012} in the waste sector; and (b) agricultural water pricing reform to improve water-use efficiency in the rice sector \cite{WaterReform_2016}. In contrast, livestock emissions intensity increases (3.3\% yr$^{-1}$), despite no clear change in year-end ruminant stocks. This rise may reflect shifts in animal farming toward a higher cattle share, driven by growing domestic demand for milk and beef (SI Table \ref{tab:ag_products}) \cite{ProvinceStats-2019-2024}. Further policy and activity details are provided in SI \ref{sec:SI_policy}.

\subsection{North China accounts for the largest emissions but shows decline in coal regions}\label{subsec3}

North and Northeast China account for over half of the anthropogenic methane emissions in the study area and also experience the strongest emissions growth. We categorize the study region into six subregions (Figure \ref{fig3}C) and estimate the average anthropogenic emissions by subregion across years: North (16.8 Tg yr$^{-1}$), Northeast (8.9 Tg yr$^{-1}$), Southwest (6.9 Tg yr$^{-1}$), Central (6.9 Tg yr$^{-1}$), East (4.9 Tg yr$^{-1}$), and South (3.1 Tg yr$^{-1}$). We identify areas with increased or decreased emissions by comparing the spatial distribution of ensemble mean between two periods: Period 1 (2019--2021) and Period 2 (2022--2024), as displayed in Figure \ref{fig3}A--\ref{fig3}B. In Period 1, inverse model estimates suggest that inventories likely overestimate emissions in the North, East, Central, and South regions, with results lower than inventories based on our regional study (by 21--45\%) and the global study by Pendergrass et al. (by 12--56\%), as illustrated in the top panel of Figure \ref{fig3}D \cite{pendergrass2025trends}. Our estimates also differ from emission inventories and results from Pendergrass et al. in some other regions: we find higher emissions in the Northeast region
, which is consistent with Liang et al.'s findings from separate regional inversions using GOSAT and TROPOMI observations \cite{pendergrass2025trends, liang_east_2022}. \\

Compared to Period 1, nearly all regions exhibit a positive growth in Period 2, except for the Central region (-0.50\ Tg yr$^{-1}$). North (+0.57\ Tg yr$^{-1}$) and Northeast (+0.79\ Tg yr$^{-1}$) regions exhibit comparatively higher emissions increases than other regions (+0.06$\sim$0.19\ Tg yr$^{-1}$), as shown in Figure \ref{fig3}C--\ref{fig3}D. Overall, our heterogeneous regional growth patterns can add nuance to inventories, which assume positive growth across all regions. Our results also indicate a higher growth in nearly all regions except the Southwest and Central, compared to results from Pendergrass et al. \cite{pendergrass2025trends}. \\ 

North China dominates coal methane emissions, but the coal-mining agglomeration shows declining emissions from Period 1 to Period 2, despite surging coal production. We identify eleven coal-mining cities in North China, each producing over 100 Mton y$^{-1}$ of raw coal and together contributing 60\% of national production during 2019--2023 (Figure \ref{fig3}A--\ref{fig3}B) \cite{NBS-coal-monthly, ShanxiYB-2023, InnerMongoliaYB-2023, YulinYB-2024}. Most major coal-mining cities show either stable or declining methane emissions, except two in Qinshui basin --- Changzhi and Jincheng in southeast Shanxi (Figure \ref{fig3}B). In addition to national efforts (Section \ref{subsec2}), reduced emissions in major coal-mining cities could have linked to two local policies: (a) mandatory pre-drainage and enhanced CMM utilization for cascading end use (urban gas, power generation, or heating) in Shanxi, which targets an enhanced utilization rate to 50\% in 2025 (up from 44\% in 2019) \cite{SX_MineralPlan_2023, CMMutilizationSX_unece_2024, liu2025methane}; (b) voluntary coal capacity restructuring, as North China retired all aging coal mines below 600 kt/y (twice the national retirement requirement) by 2021 meanwhile expanding large-scale, low-emitting mines with subsidies \cite{NRDC_CoalMineDisposal_2019, SX_CoalMineDisposal_2021}. Despite CMM utilization, the growing emissions observed in Changzhi and Jincheng might associate with the increasing mine depths over time, along with intensified CMM release rates \cite{peng2023high}. A recent study by Han et al. also indicates that the Qinshui basin accounts for 65\% of CMM emissions from major episodic events within Shanxi, largely via ventilation of underground mines, according to a satellite-based methane enhancement analysis (late 2019--early 2023) \cite{han_unveilingmethanehotspot_2024}. \\

We also find urban-related sources (0.7 Tg yr$^{-2}$, 50\%) are associated with rising methane emissions in North and Northeast China, in addition to livestock farming (0.3 Tg yr$^{-2}$, 20\%); percentages show growth share (bottom panel of Figure \ref{fig3}D). \\ 


In the North and Northeast regions, rising urban-related methane emissions could be related to increased urban gas use rather than waste infrastructure. Relative to Period 1, both regions improved waste management practices, lowering methane intensity as municipal solid waste incineration rates rose by 21\% and 39\% respectively, while maintaining very high treatment coverage for wastewater (97\%, 98\%) and sludge disposal (97\%, 99\%) (SI Figure \ref{fig:SI_urban}). In contrast, urban natural gas supply expanded toward end uses that require more last-mile gas distribution network in both regions, potentially increasing urban gas leaks. In Period 2, households and central heating together accounted for 68\% and 78\% of the incremental gas consumption (both up by 9\%) in the North and Northeast, considerably higher than their original combined shares of 24\% and 40\% in Period 1, with other uses allocated in industry and vehicles (SI Figure \ref{fig:SI_urban}). Besides pipeline leakage, marginal methane leaks can occur from residential gas appliances, including stoves, tankless water heaters, and community heating boilers \cite{zhang2022methaneNG_boilers, sun2024measurement_NG_heater_stove, xu2026methane_gasstove}. Both North and Northeast regions continue shifting from distributed coal-fired boilers and household stoves to gas-fired systems under the Clean Winter Heating Plan and the broader Coal-to-Gas Policy \cite{china_northern_clean_heating_2017, caep_2021_scatter_coal_control}. Our findings echo the GOSAT-based inverse model by Wang et al. (2022), which reports increased methane emissions conincident with rising natural gas demand in the same area during 2010--2018 \cite{wang2022atmospheric_ng_use}.

\subsection{Methane emissions in China are highly seasonal}
\label{subsec4}

Our inverse estimates imply a summer emissions peak of about 7 Tg per month in Eastern \& Central China, generally consistent with the peak time and peak magnitude of emission inventories (Figure \ref{fig4}A). The strong seasonality is mainly driven by rice cultivation activities nationwide (Figure \ref{fig4}B). Farmers typically water semi-aquatic rice fields by maintaining shallow flood levels (5--10 cm depth) to control weeds and pests, starting 1--4 weeks after sowing and ending 1--2 weeks before harvest (e.g., \cite{Bouman2007}). While this managed flooding boosts yields, it also creates an anaerobic environment that fosters microbial growth and emits methane. Middle-season rice is the dominant rice crop in China, accounting for 72\% of total yields and 67\% of rice cultivation areas within our study region, followed by double-season and early-season rice with comparable shares during 2019--2024 \cite{ProvinceStats-2019-2024}. To assess the seasonal emissions pattern, we also cross-check provincial rice cropping calendars for various crop types and derive production-weighted dates for transplantation and harvest (SI Figure \ref{fig:SI_rice}). Rice fields emit methane as early as March and April, consistent with the early-season rice transplantation month, while the emissions climb and peak in summer when the middle-season rice harvests begin in July and August \cite{li2024chinaricecalendar}. Rice crop emissions also become dominant methane sources in the Middle and Lower Yangtze Plain (East and Central regions) and the Northeast Plain (Northeast region) (Figure \ref{fig4}C). \\

Our seasonality shows a slightly different summer peak from the one inferred by Feng et al., who use a single year of TROPOMI observations from 2022 \cite{feng_2025_inverseChina}. A record flash drought hit the Yangtze River basin from June to November 2022, severely damaging early-season rice harvests in the middle--lower reaches and likely altering the seasonal variability of methane emissions under drought conditions (e.g., \cite{cao2024strongagriresilience}). Our estimate smooths out the influence of extreme weather events (i.e., droughts, floods, or extreme rainfall) by averaging monthly emissions over six years in China. \\

\section{Conclusions}
\label{sec3}

China experienced a milder methane emissions growth during years 2019--2024 than in the previous decade, contemporaneous with sustained policies for mitigation. For example, we find evidence of decoupled methane emissions from energy production, reflected in the declining CMM emissions intensity. If this decline in emissions intensity was caused by recent policies, it would show that China's methane mandates and incentives can be effective when coordinated across central and local administrative levels. If the latest CMM mitigation policies are implemented effectively, we might expect a further decoupling trend of methane emissions from coal production activities. Since 2025, China regulated emissions from coal mines with low-concentration gas (8--30\%) for the first time, and improved subsidy reward factor for CMM/CBM utilization from 1.2 to 1.5 \cite{CMMemissionstandard_2024, 2025specialfund_unconventionalgas}. The country also turned to a market-based mechanism, the trading market via China Certified Emission Reduction (CCER) program, for additional, voluntary mitigation from very low-concentration mines ($\leq$ 8\%) and ventilation air methane \cite{MEE_LowConMineGuidlines_2025}. \\

Comparatively, livestock accounts most for the recent emissions rise in our study region, and agricultural emissions intensity does not decrease as expected by the government \cite{methaneactionplan_2023}. This result indicates a possible opportunity to control emissions from enteric fermentation as part of future emissions mitigation efforts, since China has already re-utilized 80\% of animal manure mainly as fertilizer in 2025 (a goal in the \textit{Action Plan}) \cite{ManureManagement_2025}. \\

Urban gas use emerges as another rising methane source in North and Northeast China, because of stringent air quality controls and accelerated clean energy transition. It is challenging to explicitly separate urban gas sources from waste-related sources from our country-level inversion at 55-km resolution. Recently, researchers use \ce{CH_4}:\ce{C_2 H_6} ratio as a proxy to better attribute urban natural gas leaks within cities, as ethane is co-emitted with methane from natural gas but absent from waste and wastewater sources \cite{sargent2021majorityurbanNG, plant2019urbancenters, vogel2024groundmobileurban}. Additional tracers, including \ce{CO} and \ce{CO2}, are also useful in delineating the urban boundary associated with methane emissions \cite{plant2019urbancenters, whiting2026space}. However, limited surface observations of non-\ce{CO_2} species in China pose challenges to multi-species inverse model analysis.\\ 

Despite slower growth, it is unlikely that China's methane emissions have peaked. Marginal abatement costs are likely to rise over time, as low-cost mitigation options in the coal sector are increasingly exhausted \cite{khanna_assessment_2024}. Sustaining momentum will therefore likely depend on exploiting near-term methane mitigation opportunities in non-coal sectors. This challenge arguably necessitates more rigorous quantification of anthropogenic methane sources beyond coal mining through inverse models and a more established monitoring, verification, and reporting system. \\

\section{Methodology: Modeling and data}
\label{sec4}

We develop a regional inverse model to infer emissions using TROPOMI satellite observations during years 2019--2024 \cite{miller2020geostatistical}. The model is over a 0.5$^{\circ}\times$0.625$^{\circ}$ latitude-longitude grid for China, using both forward and adjoint process from a meteorology-driven atmospheric chemical transport model named GEOS-Chem. Our model enables the quantification of daily grid-box emissions, though it is challenging to provide direct uncertainty bounds on the resulting grid-level emissions estimate. Instead, we explore uncertainty using an ensemble of simulations with two sets of anthropogenic inventory under three various boundary conditions, treating the spread of inverse estimates as a proxy of the uncertainty introduced by the modeling choices. \\

We use a numerical optimization approach that iteratively couples the forward and adjoint GEOS-Chem processes to obtain posterior estimates, converging within 50 iterations \cite{chen2025spatial}. To reduce computational complexity, we execute the model separately for each year, beginning in December of the previous year for a warm-up period. Model flow diagram is in SI Figure \ref{fig:gim}. \\
 

\textbf{TROPOMI observations:} 
The TROPOspheric Monitoring Instrument (TROPOMI) was launched on the Sentinel 5 Precursor satellite in 2017, developed by the Netherlands and the European Space Agency (ESA) \cite{TROPOMI_manual2024}. It collects daily global observations of several trace gases through $\mathrm{5.5 \times 7 km^{2}}$ (nadir) pixels. Scientists at ESA retrieve TROPOMI \ce{XCH_4} measurements using a full-physics approach. This method quantifies methane absorption of sunlight reflected from the Earth's surface and atmosphere, together with physical scattering properties, based on near-infrared (NIR) and SWIR spectral bands \cite{TROPOMI_manual2024}. Known retrieval biases occur under conditions with coarse aerosol particles, high SWIR surface albedo, and varying across-track pixel indices (e.g., \cite{balasus_blended_2023}). \\

We use the corrected dry-column methane mixing ratios (\ce{XCH_4}) from the blended TROPOMI-GOSAT data, a modified TROPOMI level-2 product with lower retrieval biases \cite{balasus_blended_2023}. As quality control, we exclude observations (a) if the blended product correction exceeds 50 ppb relative to the original TROPOMI observation, as such large corrections suggest greater retrieval uncertainty; or (b) if the surface altitude is above 2 km, to limit stratospheric influence on the column and to reduce transport uncertainty over mountainous terrain at 0.5$^\circ \times$0.625$^\circ$ resolution in GEOS-Chem. The number of methane observations used is $\sim$10.8 million yr$^{-1}$ within the modeling domain. Over the domain, the annual mean difference in the dry-column mixing ratio is -91 to 74 ppb relative to the global background (Mauna Loa) and -55 to 102 ppb relative to the regional background (Mt Waliguan) (SI Figure \ref{fig:SI-tropomi_observations}). \\

\textbf{Inverse model:} The objective of our inverse model is to derive emission estimates $\boldsymbol{s}$ that match the simulated results with satellite methane mixing-ratio observations $\boldsymbol{z}$ through Bayesian statistical inference, given the spatio-temporal pattern from inventories \cite{michalak2004geostatistical, miller2020geostatistical}. $\boldsymbol{s}$ is a vector with dimensions $m\times1$, where $m$ is the number of model grid boxes at different locations and on different days. $\boldsymbol{z}$ is a vector with dimensions $n\times1$, where $n$ is the number of observations. We map $\boldsymbol{s}$ to $\boldsymbol{z}$, using $h(\cdot)$ to represent the forward operator of GEOS-Chem transport (Equation \ref{eq1}). Yet, it is impossible to perfectly match all observations because of errors in the TROPOMI observation retrievals and in GEOS-Chem model, denoted as $\boldsymbol{\epsilon}$ (dimensions $n\times1$).

\begin{equation}
\boldsymbol{z} = h(\boldsymbol{s}) + \boldsymbol{\epsilon} \label{eq1}
\end{equation}

We structure our estimated emissions $\boldsymbol{s}$ into deterministic and stochastic components, following the principles of geostatistical inverse modeling (GIM) (Equation \ref{eq2}) \cite{michalak2004geostatistical, miller2020geostatistical}. The deterministic component usually consists of a matrix of $p$ predictor variables, denoted $\mathbf{X}$ ($m \times p$), and unknown coefficients that scale the overall magnitude of each predictor $\boldsymbol{\beta}$ ($p \times 1$), to help describe known patterns in the unknown $\boldsymbol{s}$. In our model set-up, we sum the emission inventories from different sectors and use this sum as a predictor variable. As a result, $\mathbf{X}$ has dimensions $m \times 1$ in this setup, and the unknown scaling factor ($\boldsymbol{\beta}$, $1 \times 1$) will correct the overall magnitude of these inventories to be more consistent with the TROPOMI observations. Even after scaling the inventory estimates, we are unlikely to match TROPOMI observations within expected model-data errors. The inverse model therefore includes an additional term known as the stochastic component ($\boldsymbol{\zeta}$, dimensions $m \times1$) to correct the estimate $\boldsymbol{s}$ in each grid box and on each day, improving agreement with TROPOMI observations.

\begin{equation}
\boldsymbol{s} = \mathbf{X}\boldsymbol{\beta} + \boldsymbol{\zeta} \label{eq2}
\end{equation}

The optimal estimates from an inverse model typically minimize discrepancies in two aspects: (1) the divergence from prior information; and (2) the mismatch between simulated mixing ratios and observations. To solve the problem, we apply a gradient-driven algorithm that iteratively alternates between the forward process with updated emissions based on the gradients computed from the previous adjoint processes. Additional details of our inverse model are provided in the SI \ref{sec:SI_inversemodel}. \\

\textbf{Chemical-transport model:} We implement GEOS-Chem forward (v10-01b) and adjoint (v36) simulations over $0.5^\circ \times 0.625^\circ$ across 47 vertical pressure layers \cite{philip2016-GEOSChemv10, dedoussi2024-GEOSChemAdjointv36}. GEOS-Chem is a 3-D atmospheric chemistry simulation model driven by meteorological data from the Goddard Earth Observing System (GEOS) by NASA \cite{bey2001forward}. We use MERRA-2 product for meteorological inputs \cite{gelaro2017merra2}. Our modeling domain is [16.5$^{\circ}$N--55$^{\circ}$N, 71.875$^{\circ}$E--136.875$^{\circ}$E], smaller than the nested China/Southeast Asia domain in GEOS-Chem, to ensure the most efficient computation while covering the mainland of China. \\

Initial conditions and boundary conditions are two key parameter settings related to mixing ratios in the GEOS-Chem model. In each annual run, we set initial conditions by interpolating December methane mixing ratios from the previous year based on CAMS global optimized greenhouse gas inversions product ($1^{\circ}\times 1^{\circ}$, 34 vertical layers) \cite{CAMS_description_2025}. \\

\textbf{Boundary conditions:} We use outputs from three separate global models to construct the sensitivity for boundary conditions. The first data source is the CAMS global greenhouse gas reanalysis (EGG4) product, which provides atmospheric methane mixing ratios ($0.75^{\circ}\times 0.75^{\circ}$, 25 layers) \cite{agusti2023CAMS-EGG4}. We interpolate this product on a $2^{\circ}\times 2.5^{\circ}$ horizontal grid with 47 vertical levels. The other two sources are simulated mixing ratios from in-house JHU global inverse modeling simulations that assimilate the original TROPOMI and blended-TROPOMI products, respectively, both implemented with GEOS-Chem ($2^{\circ}\times 2.5^{\circ}$) \cite{chen2025spatial}. For all three sources, we further apply a temporally splined correction at the four boundaries of the regional model domain, to more accurately represent the seasonal variability of methane inflows and outflows across the model domain. More details on aligning boundary conditions are in SI \ref{sec:SI_bc}. \\

\textbf{Emissions inventories in the inverse modeling prior:} We use a combination of regional (when available) or global inventories for each sector. These inventories are the default inventories used in the Harmonized Emissions Component (HEMCO) version 3.10.1, a software that re-grids multiple inventories for atmospheric modeling applications \cite{lin2021HEMCO3}. The inventories include both anthropogenic and natural sources. \\

For anthropogenic sources, we have two different input configurations: (a) In one configuration, referred to ``EDGARv8'', we use EDGAR v8.0 for all anthropogenic sectors; (b) In another configuration, reffered to ``Composite'', we rely on EDGAR v8.0 to represent most sectors, including urban waste disposal, wastewater, livestock, and other activities \cite{crippa_insights_2024}. Yet, for coal, oil, and gas sectors, we instead adopt the Global Fuel Exploitation Inventory version 3 (GFEI v3), which provides updated, localized coal emission estimates for China \cite{scarpelli2025GFEIv3}. We also use the Global Rice Paddy Inventory (GRPI) for rice paddy emissions in China, as it offers more recent revisions for rice-related emissions inferred from the inundation map in 2022 \cite{chen2025GRPI}. \\

For natural sources and sinks, we use Global 0.5-deg Wetland Methane Emissions and Uncertainty (WetCHARTS version 1.3.1) to represent wetlands \cite{bloom2017globalWetCHARTS}, Global Fire Emissions Database version 4 (GFED v4) for biomass burning, and a range of dedicated inventory datasets for specific processes in default settings: geological seeps \cite{etiope2019geologicseeps}, hydroelectric reservoirs \cite{delwiche2022reservoir}, termites \cite{fung1991termites}, and soil uptake \cite{murguia2018soil}. \\

\textbf{Sector attribution and trend analysis:} We estimate sector-specific emissions in two steps, following an approach similarly applied in previous studies (e.g., \cite{he2025attributing, zhang_observed_2022, chen_methane_2022}). First, for each grid box, we calculate the share of each sectoral source relative to the total emissions from all sectors, based on the compiled HEMCO emission inventories. These sectoral fractions vary by location and day. Next, we multiply posterior estimates with these corresponding sectoral shares to obtain posterior sectoral emissions for every grid box. This attribution approach does not rely on overall posterior estimates, but can bring uncertainty when inventories over-/under-estimate the relative magnitude of sectoral sources in some areas or days. For additional analysis, we compare sectoral emission trends against the associated production activities over time, and derive the resulting emissions intensities. We focus on the primary production processes that are responsible for sectoral methane emissions: raw coal production for coal mining, crude oil and natural gas production for oil and gas extraction, year-end stock of ruminant animals (including cattle, sheep, goat, and camel) for livestock, rice crop yields for rice cultivation, as well as garbage and wastewater discharged for urban waste. All activity data is at the provincial level and obtained from government sources, including National Bureau of Statistics \cite{ProvinceStats-2019-2024} and China Urban Construction Year Book \cite{UrbanConstructionYB2019-2024}. \\

\textbf{Inferring urban sources:} We note that urban sources are usually spatially co-located within the same model grid box (e.g., landfills, wastewater treatment plants, gas distribution and use), making it difficult to disaggregate urban sources using the inverse modeling results. Hence, we analyze urban gas use activity and waste infrastructure datasets for both regions to infer the possible contribution of different sources. \\

\textbf{Study region:} We discuss findings about Eastern \& Central China (see divisions in SI Figure \ref{fig:regioncat}), due to suspected observation retrieval biases and low signal-to-noise ratio in the western regions. TROPOMI retrievals are very sensitive to either too high or too low short-wavelength infrared (SWIR) surface albedo (e.g., \cite{balasus_blended_2023}). Western China, especially the northwestern regions and the Tibeten Plateau, likely exhibits confounding bias in TROPOMI observations due to the complex terrain: extensive deserts (with high SWIR surface albedo) and snow-covered mountain ranges (with low SWIR surface albedo) \cite{balasus_blended_2023, alles2013china}. A recent inverse modeling study also reports higher estimated methane emissions from northwestern East Asia (including northwestern China) using TROPOMI than Greenhouse gases Observing SATellite (GOSAT), likely due to greater seasonal TROPOMI retrieval biases in the region during December to March \cite{liang_east_2022}. In addition, Western China accounts for only 8\% of total emissions in the country with sparse emission signals, according to inventories. As a result of these confounding biases, we focus our analysis on the eastern and central regions of China without compromising the quality of the conclusions. \\

\textbf{Configuration:} We configure GEOS-Chem models and carry out the inversion experiments at the Advanced Research Computing at Hopkins (ARCH) core facility (rockfish.jhu.edu).


\newpage
\begin{landscape}

\section{Figures}\label{sec5}

\begin{figure}[htb!]
\centering
 \includegraphics[width=0.95\linewidth]{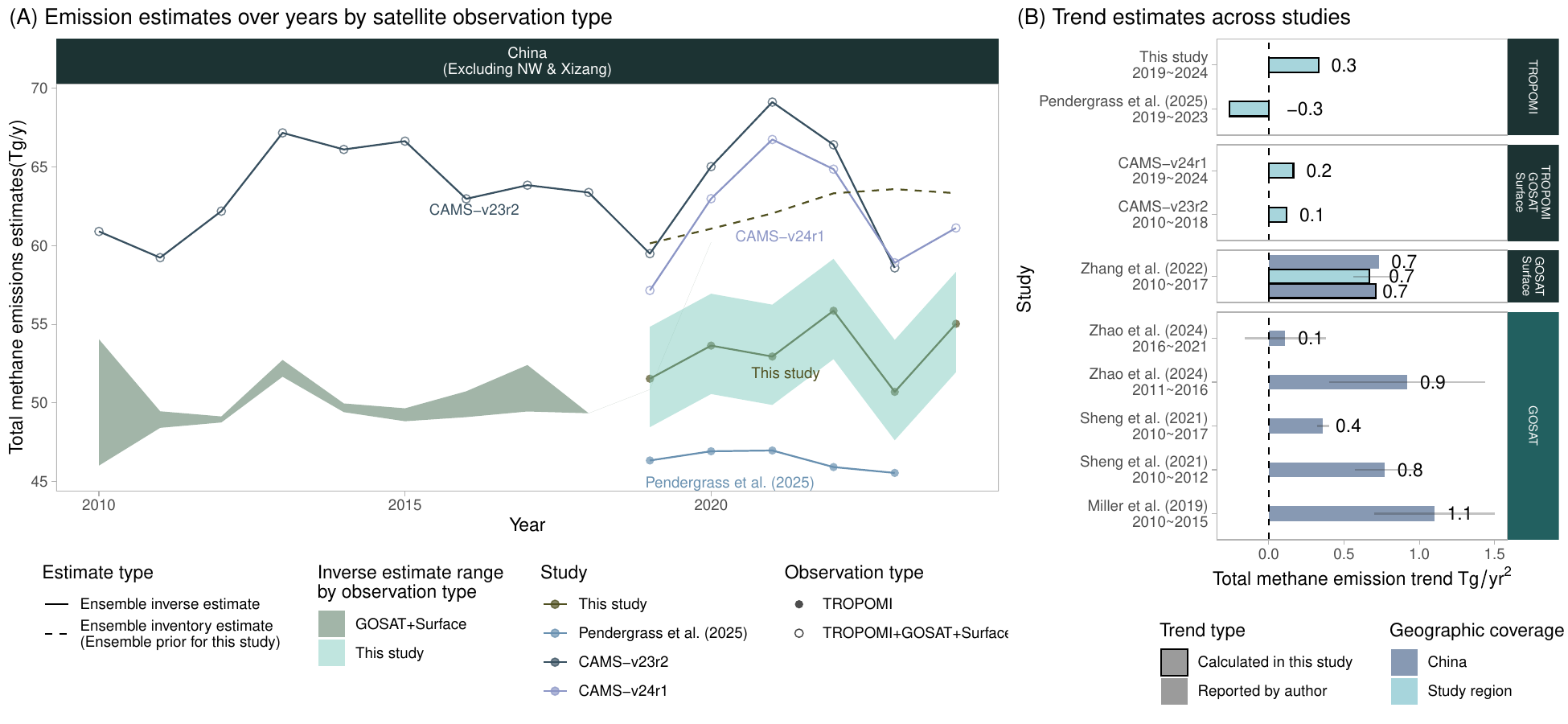}
 \caption{\textbf{Cross-study comparison of \ce{CH4} emissions and trend estimates from China: } (A) Comparison of posterior emission estimate ranges in Eastern \& Central China over years between our estimates and the ensemble mean of existing satellite-based inversion studies \cite{CAMS_description_2025, pendergrass2025trends, CAMS_evaluation_2022, zhang_observed_2022}. The annual variations from ensemble mean across estimates range from -3.08 Tg y$^{-1}$ to 3.30 Tg y$^{-1}$, which is also treated as the uncertainty of estimates for all years; (B) Trend estimates from existing satellite-based inversion studies by observation type, such as TROPOMI, GOSAT, GOSAT+Surface, and TROPOMI+GOSAT+Surface \cite{CAMS_description_2025, pendergrass2025trends, CAMS_evaluation_2022, zhang_observed_2022, miller_chinas_2019, sheng_sustained_2021, zhao_slowdown_2024}. We estimate trends for study regions when output files are available from other studies; otherwise, we use nationwide trend values reported by the authors for comparison. Both panels include regional and global studies that cover China for comparison. For global models, CAMS optimizes daily fluxes at the $1^{\circ} \times 1^{\circ}$ resolution, while Pendergrass et al. quantify monthly scaling factors for clustered emissions at $2^{\circ} \times 2.5^{\circ}$ resolution. }
 \label{fig1}
\end{figure}

\clearpage

\begin{figure} [htb!]
    \centering
    \includegraphics[width=1\linewidth]{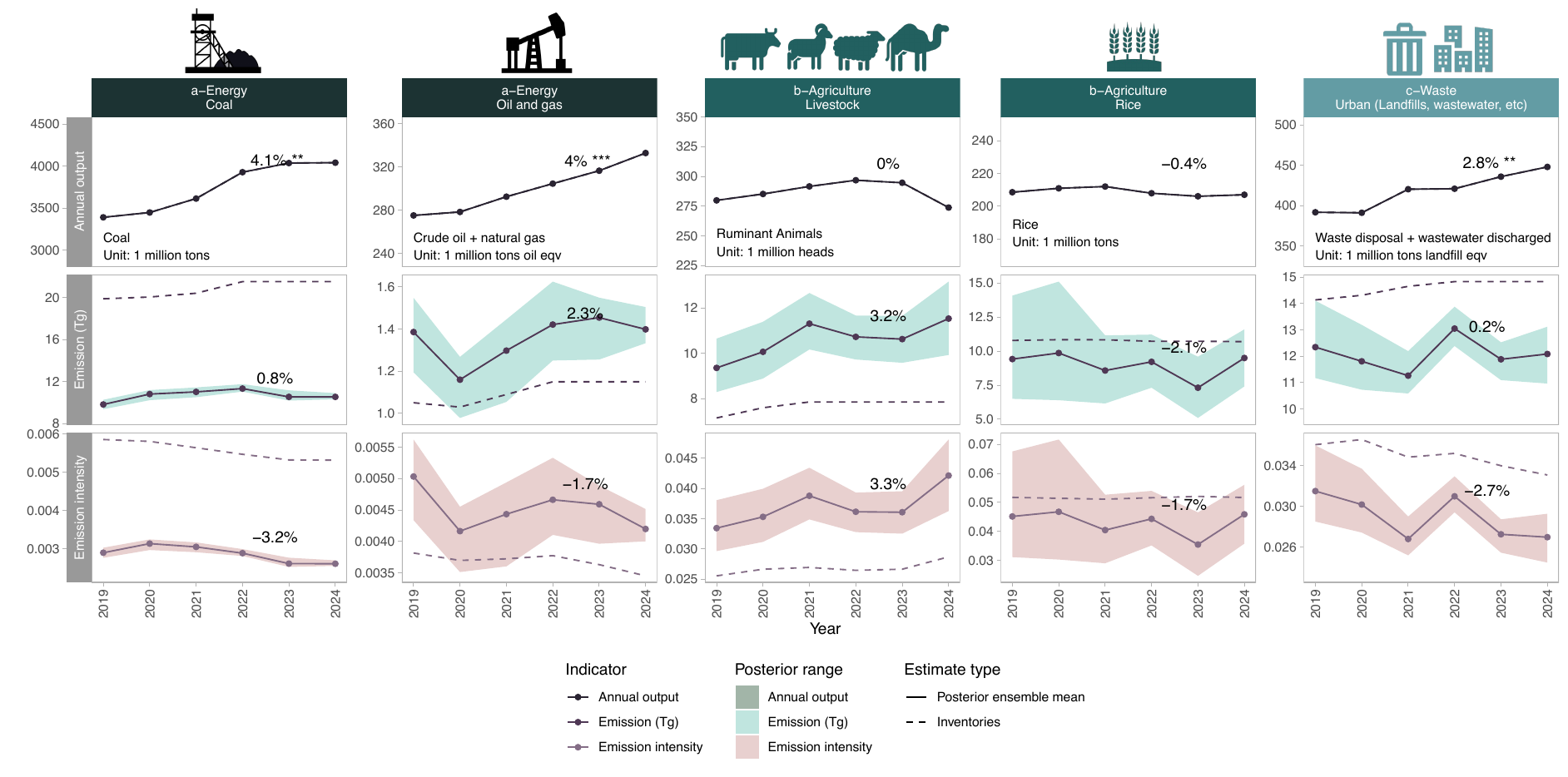}
    \caption{\textbf{Sectoral production activities, methane emissions, and emissions intensity over years in Eastern \& Central China: } The plot shows the sectoral production activities (top row), associated anthropogenic methane emissions (middle row), and methane emissions intensity (bottom row) over time. Emissions intensity is the ratio of emissions to production in units. Value above the line in each panel represents the linear average annual growth rate from 2019 to 2024, with ***, **, and * indicating p values below 0.001, 0.01, and 0.05 respectively. 
    The spread shows the uncertainty from varying modeling choices in boundary conditions and inventories for prior estimates. We aggregate the production activities using annual provincial production data from China Urban Construction Year Book (for garbage and wastewater discharged) \cite{UrbanConstructionYB2019-2024} and National Bureau of Statistics (for all other sectors) \cite{ProvinceStats-2019-2024}. To combine some sub-sectors, we use the following conversion rates: (1) 0.86 million tonnes of oil equilvalent per 1 billion cubic meter of natural gas, based on BP conversion \cite{bp_conversion_2022}; (2) 3 million tonne of landfill equivalent per billion m$^3$ of wastewater in terms of methane emissions (given 0.25 kg biochemical oxegen demand (BOD)/m$^3$ wastewater \cite{gerba2009wastewater}, 0.6 kg \ce{CH_4}/kg BOD \cite{IPCC2006_Waste_Vol5}; and 0.05 kg\ce{CH_4}/kg solid waste \cite{IPCC2006_Waste_Vol5}).
    }
    \label{fig2}
\end{figure}


\end{landscape}

\clearpage

\begin{figure}[htb!]
    \centering
    \includegraphics[width=0.95\linewidth]{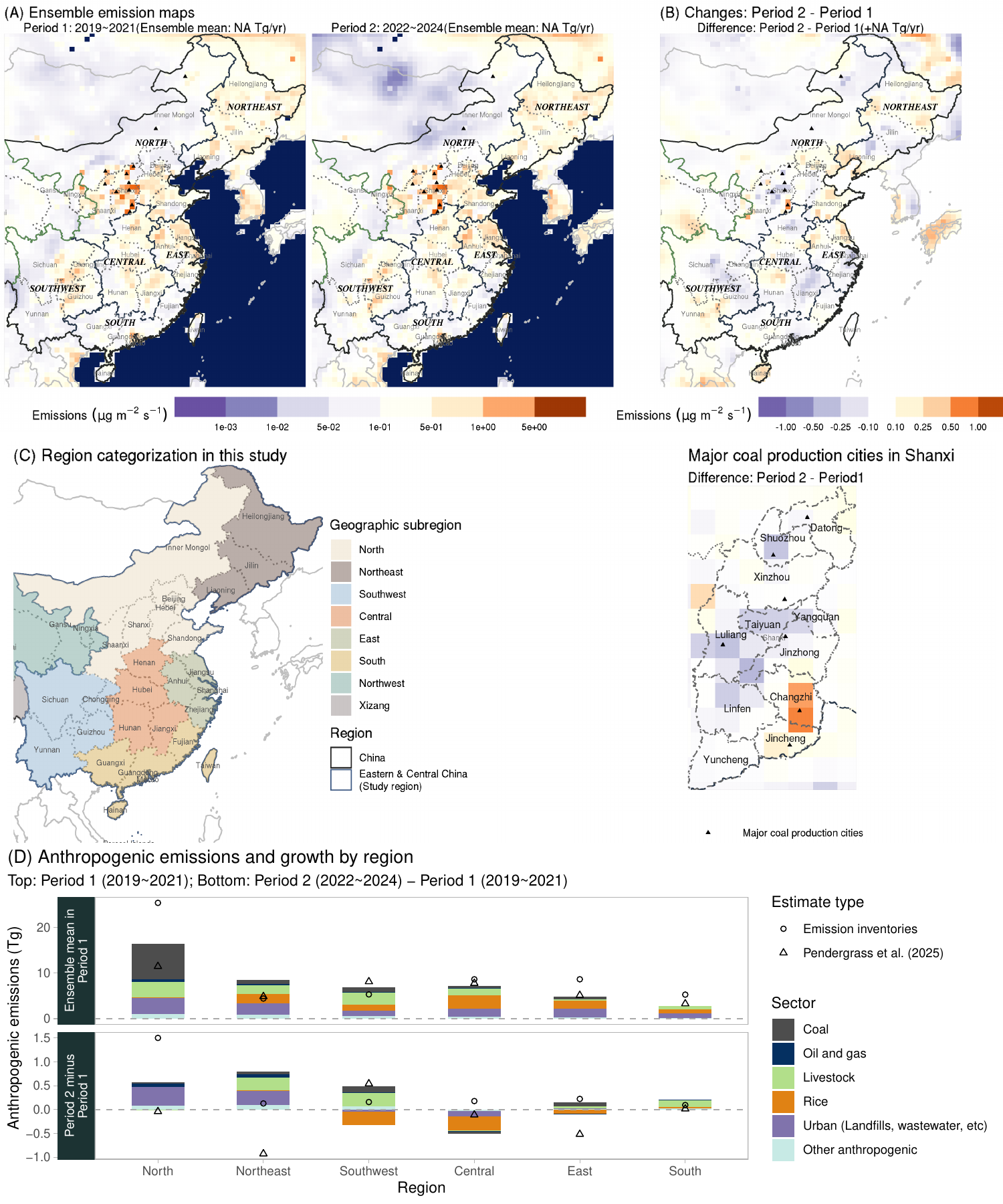}
    \caption{\textbf{Spatial distribution of \ce{CH4} emissions and changes in Eastern \& Central China: }
    (A) Ensemble mean emissions map during the 1st half period (Period 1: 2019-2021) and the 2nd half period (Period 2: 2022-2024); The color scale is based on a logarithmic scale.
    (B) Emissions change between Period 2 and Period 1 of the study region (top) and Shanxi province (bottom); 
    (C) Coverage of each subregion (North, Northeast, Southwest, Central, East, and South);
    (D) Methane emission sources in Period 1 and the changes in Period 2 (relative to Period 1) for each subregion. 
    In Panel A--B, black triangles mark the cities that produce the most raw coal within the region during 2019--2023, located in Shanxi (Shuozhou, Changzhi, Datong, Lyuliang, Jincheng, Xinzhou, and Jinzhong), Shaanxi (Yulin), and Inner Mongolia (Ordos, Xinlinguole, and Hulunbeier) (SI Table \ref{tab:coal_cities}). Rectangle labels represent in-situ surface towers from \ce{CH_4} GLOBALVIEWplus v7.0 \cite{Schuldt2024} and NRT v7.1 ObsPack \cite{Schuldt2025}.
    }
    \label{fig3}
\end{figure}

\newpage
\begin{figure}[htb!]
    \centering
    \includegraphics[width=1\linewidth]{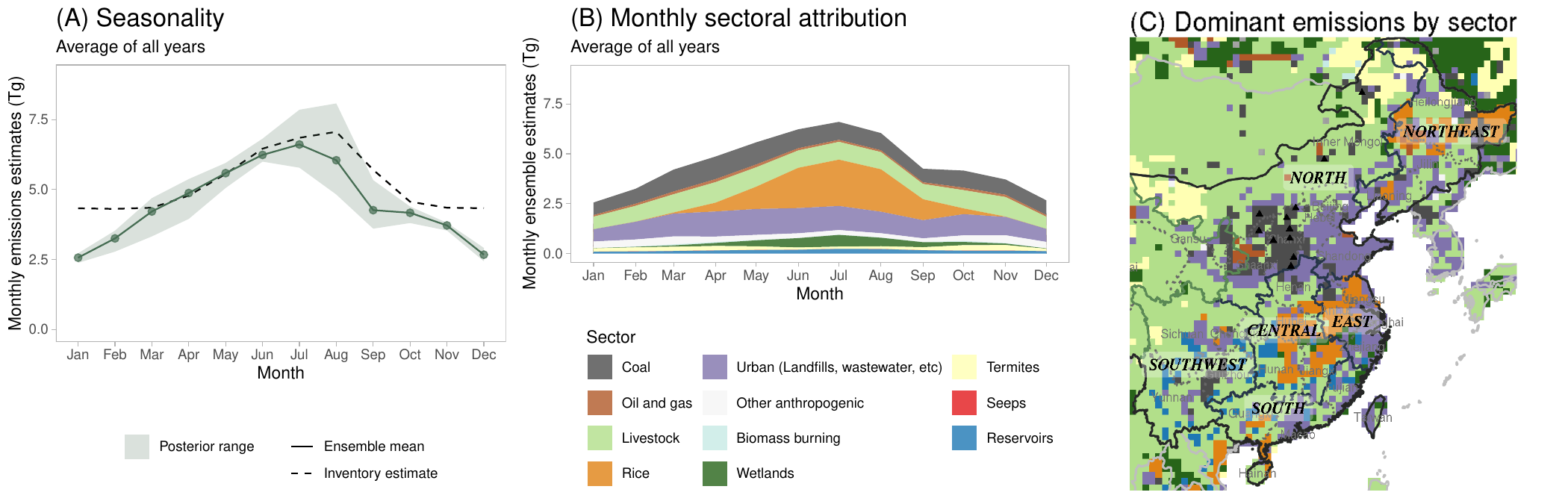}
    \caption{\textbf{\ce{CH4} emissions seasonality and sectoral attribution in Eastern \& Central China: }
    (A) Comparison of weighted monthly emissions from inverse estimates (posterior) and emissions inventories (prior), where we apply weights to normalize all months to 30 days;   
    (B) Composition of the ensemble mean seasonality with sectoral attribution;
    (C) Sectoral dominance at the grid cell level (in terms of annual emissions).}
    \label{fig4}
\end{figure}

\backmatter
\bmhead{Supplementary Information} Please check out the supplementary information document below. \\

\bmhead{Acknowledgements} We thank the GEOS-Chem Support Team for their work in developing and sustaining the GEOS-Chem community. We also acknowledge the European Space Agency (ESA) and the Copernicus Programme for supplying the Sentinel-5P TROPOMI data. Additionally, we are grateful for the insightful comments received from the Atmosphere \& Ocean Seminar at Johns Hopkins University (led by Dr. Darryn Waugh and Dr. Anand Gnanadesikan) and from participants at the American Geophysical Union Annual Meetings (2024, 2025).

\section*{Declarations}

\bmhead{Funding} This work is funded by an NSF CAREER award (\#2237404) and a Johns Hopkins Catalyst Award. Ziting Huang also thanks the financial support from Caplan Scholarship Fund in Climate Change and the Environment offered by Johns Hopkins University. 



\bmhead{Data availability} 
We obtained the blended TROPOMI+GOSAT satellite atmospheric methane data product (by the Harvard Atmospheric Chemistry Modeling Group) from \url{https://registry.opendata.aws/blended-tropomi-gosat-methane/}. The MERRA-2 dataset, produced by the Global Modeling and Assimilation Office (GMAO) at NASA’s Goddard Space Flight Center, was downloaded from \url{http://geoschemdata.wustl.edu/} for application in the GEOS-Chem model. We acquired CAMS-related datasets from the Atmosphere Data Store: CAMS global greenhouse gas reanalysis EGG4 product (\url{https://ads.atmosphere.copernicus.eu/datasets/cams-global-ghg-reanalysis-egg4}) and CAMS global inversion-optimised greenhouse gas fluxes and concentrations (\url{https://ads.atmosphere.copernicus.eu/datasets/cams-global-greenhouse-gas-inversion}). In addition, we collected yearbook data from the National Bureau of Statistics of China (\url{https://www.stats.gov.cn/english/}) and from the China Statistical Yearbooks Database provided by the China National Knowledge Infrastructure (CNKI).

\bmhead{Materials availability}
The Goddard Earth Observing System - Chemistry (GEOS-Chem) model is a chemical transport model available at \url{https://geoschem.github.io/index.html}. The Harmonized Emissions Component (HEMCO) model is included within the GEOS-Chem framework and is also publicly accessible via \url{https://hemco.readthedocs.io/en/stable/}.

\bmhead{Code availability}
Code of inverse modeling is available at \url{https://github.com/greenhousegaslab/geostatistical_inverse_modeling.git}.

\bmhead{Artificial Intelligence (AI) Disclosure}
During the preparation of this work, the authors used Overleaf AI LaTeX tool and ChatGPT-5 in order to proofread the draft and to adjust the emotional tone of sentences. After using this tool/service, the authors reviewed and edited the content as needed, and take full responsibility for the content of the publication.

\bmhead{Author contribution}
Z.H. and S.M.M. led the study, designed the experiments, and wrote the manuscript with inputs from all co-authors. Z.H. and S.M.M. performed the inverse model, simulations, and result analysis. Z.H, S.M.M, A.C and L.F developed the regional inverse model for methane estimates, with module inputs from L.T.M. and X.Y.. Z.H. collected the production data and reviewed methane-related policies. Z.H. and S.M.M. discussed results with inputs from B.F.H., A.C., W.S.D., D.C.G., and K.L.M.. All authors commented on the manuscript.

\noindent

\newpage

\begin{appendices}



\end{appendices}

\newpage
\putbib[01-sn-bibliography]

\end{bibunit}

\clearpage

\newpage
\null
\thispagestyle{empty}
    \vspace*{\fill}
    \begin{center}
        This page intentionally left blank.
    \end{center}
    \vspace*{\fill}
\newpage

\resetlinenumber[1]

\begin{bibunit}[sn-mathphys-num]

\clearpage

\setcounter{page}{1}
\renewcommand\labelenumi{(\alph{enumi})}
\setcounter{figure}{0}
\renewcommand{\thefigure}{S\arabic{figure}}
\renewcommand{\theHfigure}{S\arabic{figure}} 
\renewcommand{\figurename}{SI Figure.}
\setcounter{table}{0}
\renewcommand{\thetable}{S\arabic{table}}
\renewcommand{\theHtable}{S\arabic{table}}
\renewcommand{\tablename}{SI Table.}

\begin{center}
    {\LARGE\bfseries Supplementary Information\par}
    \vspace{1.5em}
    {\LARGE Satellite-based emissions estimate indicates progress toward China's methane mitigation goals\par}
    \vspace{3em}
\end{center}


\begin{appendices}

\section{}

\subsection{Elaboration of non-coal sectoral policies in China} \label{sec:SI_policy}
As with the coal sector, we compare the emissions intensity trend with relevant policies and developments in the other sectors. We sort out sectoral policies that directly or indirectly affect methane emissions for the oil and gas industry, waste sector, and agriculture sector. All activity data presented here pertains to our study region, unless explicitly stated as national or provincial.
\begin{enumerate}[label=(\alph*)]
    \item \textbf{Oil and gas industry:} Our findings on the likely decline in emissions intensity are consistent with the reports published by Chinese companies. China National Petroleum Corporation (CNPC), the country's largest oil and gas company, claimed a 44\% reduction in its \ce{CH_4} emission intensity metric from 0.50\% in 2019 to 0.28\% in 2024, moving toward its 2025 target of limiting \ce{CH_4} emissions from operated chains to 0.25\% of total marketed natural gas by volume \cite{CNPCreport2024, OGCI2025}. This emissions intensity trend is concurrent with the voluntary formation of the China Oil and Gas Methane Alliance in 2021, where operators share practices in abatement technologies, leak detection, and vented gas utilization, in response to the enforced \ce{CH_4} reporting and monitoring of oil and gas fields since 2019 \cite{OGAlliance_2021, OGIndustry_2019}.
    
    \item \textbf{Waste sector:} The waste sector also shows a plausibly reduced \ce{CH_4} emission intensity trend, contemporaneous with policies on zero-waste cities, resource conservation, and the subsidized waste-to-energy (WtE) power \cite{ZeroWasteCity2021, NRDC_subsidyWtE_2012}. For urban waste, as WtE plants rapidly expanded, the share of urban domestic garbage treated by incineration rather than sanitary landfilling rose from 50\% in 2019 to 82\% in 2024 within our study region \cite{UrbanConstructionYB2019-2024}. For wastewater, actions span across sources, sewer networks, and treatment plants. New regulations require industrial users to pre-treat effluent to meet chemical oxygen demand (COD) standards and mandate real-time COD monitoring at wastewater treatment plants \cite{MEEWastewaterMonitor_2020, MEEWastewaterManagement_2020}. Local governments have also separated stormwater and sewage through sewerage upgrades to reduce overflow-related anaerobic zones, lowering the share of combined drainage pipelines by 7\% since 2019 \cite{UrbanConstructionYB2019-2024, MEELocalDischargeOutcome_2023}. Overall, wastewater treatment rates increased from 97\% to 99\%, alongside a rise in sludge disposal rates from 97\% to 100\% during 2019–-2024 \cite{UrbanConstructionYB2019-2024}. All these measures, though not mitigation-specific, seem linked to the reduced \ce{CH_4} emission intensity in the waste sector. Nationwide, China has already achieved the 2025 waste-sector goals set out in the \textit{Action Plan}: it incinerated 85\% of urban waste for power and heating ($>$60\% target) and treated nearly all dry wastewater sludge ($>$90\% target) \cite{STDaily_MethaneProgress_2025, UrbanConstructionYB2019-2024}.
    
    \item \textbf{Agricultural sector:}  The slight decrease in rice emissions intensity may be related to the 10-year agricultural water pricing reform initiated in 2016, which is designed to improve water-use efficiency \cite{WaterReform_2016}. As rice cultivation relies on flooding for weed control, the water-saving reward mechanism could motivate farmers to adopt a more efficient irrigation system, such as alternate wetting and drying (AWD), which potentially reduces 23--72\% of \ce{CH_4} emissions with minimal yield loss \cite{setyanto2018alternate, kraus2022greenhouse,loaiza2024evaluating}. Extreme weather events also lead to fluctuations in emissions intensity due to reduced rice harvest in some years, including drought, heatwaves, and extreme rainfalls during 2022--2024 \cite{chen2025projectedextremecliamterice, SCMPflooding_2024}. In the livestock sector, the increased emissions intensity might associate with the structural change of animal farming activities toward a higher cattle share. We focus on enteric fermentation here, the primary source of livestock emissions with a share of 83--86\% in China, based on EDGAR v8.0 \cite{crippa_insights_2024}. There is an increased share of cattle among total ruminant stocks between 2019--2024 (by +3\% in both year-end stock and annual average stock), along with the rising demand for milk  (+19\%) and beef (+17\%) in our study region (SI Table \ref{tab:ag_products}). Methane emissions from a single cow are equal to those emissions from 10--15 sheep \cite{broucek2014livestockmethane}. Therefore, a small rise in the proportion of cattle (1--3\%) leads to a disproportionately larger increase in methane emissions from ruminant animals.
\end{enumerate}

\subsection{Further details of inverse model} \label{sec:SI_inversemodel}

The motivation of quantifying daily grid-box emissions is to better capture local and day-to-day methane variations. For more robust analysis, we aggregate these results to monthly and annual scales. In addition, our inverse model provides sufficient flexibility in deriving these estimates through stochastic components and error covariance matrices that account for spatiotemporal correlation, which thereby reduces the overfitting of unexplained extreme values. Here, we present additional technical details of our inverse model, specifically about how to incorporate uncertainty in the model, the optimization algorithm used, the transformation that enforces non-negativity, and the validity of TROPOMI observations.  \\

\textbf{Covariance matrices:} Our inverse model accounts for two classes of uncertainties, and it does so using covariance matrices: (a) errors in emission inventories; (b) errors due to the atmospheric transport model and satellite retrievals (referred to here as model-data mismatch errors). We construct an error covariance matrix ($\mathbf{Q}$) to represent three aspects of uncertainty in the emission inventories: the standard deviation of the errors, spatial correlations in these errors, and temporal/seasonal correlations in these errors \cite{miller2020geostatistical}. For spatial or temporal error covariances, we use a spherical covariance model function to show de-correlation over time and space (e.g., \cite{Kitanidis_1997c}). We estimate parameter values for each error in \textbf{Q} based on emission inventories, before running the inverse model. In our inverse model, we set maximum correlation scales of 350 km in space and 90 days in time for $\mathbf{Q}$. This covariance matrix $\mathbf{Q}$ is used to help guide the spatial and temporal structure of the stochastic component in Function 2 \cite{miller2020geostatistical}. For model-data mismatch error, we set the diagonal elements of the covariance matrix ($\mathbf{R}$) to be $\mathrm{(25\ ppb)^{2}}$ to represent errors in the GEOS-Chem and TROPOMI observations, as suggested by another study \cite{chen2025spatial}. \\

\textbf{Gradient-based algorithm:} We use the quasi-Newton solver, named Limited-memory Broyden-Fletcher-Goldfarb-Shanno (L-BFGS), to efficiently approximate optimal estimates with the deepest gradient through forward-adjoint iterations (e.g., \cite{nocedal1980lbfgs, liu1989lbfgs}). Our numerical approximation can work in very high dimensions, as it does not rely on an explicit adjoint matrix or computational effort to get large matrices inverted. Our inverse model is different from methods used in a reduced size problem, which typically optimize annual/monthly scaling factors for spatially clustered regions (e.g., \cite{chen_methane_2022, liang_east_2022}). \\

\textbf{Non-negativity assurance:} We apply a square-root transformation on emission estimates, mainly for two reasons: (a) Non-negativity constraints on emissions estimates. We assume inverse model focuses on optimizing net positive surface fluxes, while GEOS-Chem accounts for atmospheric methane sinks via chemical oxidation. Estimates in the original space may overfit errors in the observations in some locations or days, at the cost of unrealistically negative emission values elsewhere, under the standard multivariate Gaussian assumptions in the inverse model; and (b) Higher accuracy in estimates with a transformed space closer to multivariate Gaussian distribution. In the original space, emissions are strongly right-skewed with heavy tails, where few locations or days have extremely large values. These long tails can make the joint distribution substantially deviate from a multivariate Gaussian. The square-root transformation can reduce the skewness by compressing extreme values and improve the accuracy of the inverse model. \\

\textbf{Uncertainties in observations:} We use the blended TROPOMI+GOSAT satellite data product (``blended TROPOMI observations'') with quality filters in the inverse model (see Section \ref{sec4} in main text) \cite{balasus_blended_2023}. From 2019 to 2024, the average correction applied to raw TROPOMI XCH$_4$ observations across the modeling domain is -2.0 ppb (median: -2.5 ppb; range: -21.4 to 31.4 ppb), as displayed in SI Figure \ref{fig:SI-correction_tropomi}-B. Raw TROPOMI observations are adjusted higher in most parts of the Western China while lower in Tibetan Plateau and the Northern Xinjiang (SI Figure \ref{fig:SI-correction_tropomi}-A). \\

The overall magnitude of the corrections grows marginally over time in the blended TROPOMI observations. In Period 2 (2022--2024), the average correction factor shows a slight increase across our model domain, with an average change of +1.3 ppb (median: 1.3 ppb, range: -22.8 to 20.7 ppb), as shown in SI Figure \ref{fig:SI-correction_tropomi}-B. This positive change reflects a weaker negative and a stronger positive adjustment across most of China during Period 2, compared to Period 1 (2019--2021). If the correction had remained constant over our study period, we would expect to see slightly lower blended TROPOMI observations in Period 2 than the observations currently used in the inverse model.We would then expect to see an even flatter emissions trend from the inverse model than we actually see with the time-varying TROPOMI data correction. Overall, this issue potentially adds additional uncertainty to our estimated emissions trend. However, this uncertainty is unlikely to change the overall conclusions of our paper; if the time variation in the TROPOMI correction were erroneous, it would lead us to over-estimate the emissions trend for China. In the manuscript, we argue that China's emissions have flatted relative to the previous decade, and this uncertainty in the TROPOMI data correction could yield an even flatter trend.\\

\subsection{Boundary conditions} \label{sec:SI_bc}

We adjust mixing ratios from global inverse models to serve boundary conditions for two reasons. First, our regional inverse model has a distinct set up from any of the global models, using different inventories, observation products, observation quality filters, or even a different chemical transport model. Second, the coarser-resolution model may have weaker tropospheric vertical transport at mid- to high latitudes, a known issue in the GEOS-Chem community \cite{stanevich_2020_ctm_resolution}. As a result, global inverse models could oversimplify mixing-ratio patterns over western mountainous regions, even when using the same inverse model setup with GEOS-Chem as regional models. \\

Therefore, we fit the simulated mixing ratios from global models with finer details provided by CAMS-EGG4, whose primary aim is to optimize long-term atmospheric variability instead of fluxes (Table \ref{tab:BCcorrection}) \cite{agusti2023CAMS-EGG4}. For all boundary conditions, we apply spline correction techniques, to reduce the \ce{XCH_4} mismatch pattern in time (and in space) against TROPOMI observations (Table \ref{tab:BCcorrection}). The idea is to have simulated \ce{XCH_4} results that match TROPOMI observations within GEOS-Chem buffer zones. In our GEOS-Chem setup, buffer zones consist of the outer edges of the domain (a width of three grid boxes) and serve as transition regions that only account for global influences through boundary conditions. In regional GEOS-Chem simulations, buffer zones have no advection but still reflect meteorological variations \cite{geoschem_nested_grid_guide}. Thus, we can rely on buffer zones to match impacts of boundary conditions with TROPOMI XCH$_4$ observations, as these zones are independent of flux optimization in our inverse model.\\

Despite attempts at refinement, it remains difficult to establish truly “ideal” boundary conditions for China. China has strong methane inflows associated with coal-mining, oil, and gas emissions from Central Asia in the west, Mongolia and Russia in the north, and Southeast Asia and India in the south (SI Figure \ref{fig:SI-wind_observations}). The intricate topography in the west further adds complexity to the inflow situation. Additionally, the blended TROPOMI+GOSAT product shows large adjustments in Central Asia (upward) and along the western (mixed) and northern (upward) boundaries of our China domain, indicating substantial uncertainty in TROPOMI observations in these areas \cite{balasus_blended_2023} (SI Figure \ref{fig:SI-correction_tropomi}). The limited availability of in-situ measurements also hinders accurately constraining emissions from Central Asia in global models, likely underestimating methane inflows into China. 

\begin{center}
\begin{table}[ht]
\caption{Different boundary conditions used in inversions}
\label{tab:BCcorrection}
\begin{tabular}{ cccc } 
 \hline
        No.
       & Boundary condition source \footnotemark[1]          
       & EGG4-adjusted?   \footnotemark[2]                   
       & Correction type? \footnotemark[3]      \\
 \hline
 BC1 & CAMS-EGG4                              
       & -                                 
       & Temporal              \\ 
 \hline
 BC2 & \makecell{Global inversion \\(TROPOMI)}           
       & \makecell{Yes \\(Monthly average by layer)}           
       & Temporal               \\
 \hline
 BC3 & \makecell{Global inversion \\(blended-TROPOMI)}        
       & \makecell{Yes \\ (Monthly average by layer)}           
       & Temporal               \\
 \bottomrule
 \hline
\end{tabular}
 \footnotetext{Note: All adjustments and corrections are applied separately to each domain bound (north, south, west, and east).}
 \footnotetext[1]{In run 1, we use interpolated mixing ratios from CAMS-EGG4 ($0.75^{\circ} \times 0.75^{\circ}$), which has valid data till the end of 2020. In run 2--4, we use simulated mixing ratios for all years, based on in-house global inversions with GEOS-Chem as the chemical transport model at $2^{\circ} \times 2.5^{\circ}$ resolution.}
 \footnotetext[2]{We match monthly simulated mixing ratios in run 2--3 with interpolated CAMS-EGG4 products across layers. As CAMS-EGG4 do not provide valid reference after 2020, we assume monthly discrepancy factors of each bound across layers stay consistent with 2020 for years 2021--2024.}
 \footnotetext[3]{We derive  adjusted mixing ratios ($\mathrm{\ce{CH_4_{Adjusted}}}$) in three steps: (i) We first calculate actual daily discrepancy factors ($\mathrm{f_{D}}$) for each bound, where $\mathrm{f_{D}=\frac{\ce{XCH_4_{Model}}-\ce{XCH_4_{TROPOMI}}}{\ce{XCH_4_{Model}}}}$. (ii) For temporal correction, we derive daily correction factors ($\mathrm{f_{C}}$) through the spline interpolation (DF=12) over $\mathrm{f_{D}}$ to avoid over-fitting. 
 Each bound has an individual spline interpolation. 
 (iii) We further apply a multiplier (M) on $\mathrm{f_{C}}$, linearly degraded from surface layers ($\mathrm{M=100\%}$) to top layers ($\mathrm{M=0\%}$), to reduce over-correction on stratosphere. Hence, the adjusted mixing ratios of each layer is $\mathrm{\ce{CH_4_{Adjusted}} = (1 - M\ f_{C})\  \ce{CH_4_{Model}}}$.}
  \footnotetext[4]{The ensemble simulations include six run types with modeling input choices from two anthropogenic inventory sets under three boundary conditions. These inventory combination for sources include (A) \textbf{EDGARv8:} EDGAR v8.0 for all anthropogenic sectors; (B) \textbf{Composite:} GFEI v3.0 for energy, GRPI for rice, and EDGAR v8.0 for urban, livestock, and other.}
\end{table}
\end{center}

\subsection{Model-measurement comparison}  \label{sec:SI_model-data}
We compare prior and posterior simulation results against measurements, including TROPOMI XCH$_4$ and in-situ network measurements (SI Figure \ref{fig:SI-insitu_measurements}). Posterior simulations show an enhanced performance in aligning TROPOMI XCH$_4$ measurements, with R$^{2}$ values increased by 16\%--24\% and RMSE values decreased by 30\%, relative to prior simulations. In contrast, results of prior and posterior simulations show comparable correlations with the in-situ observations. Similarly, the analysis by Liang et al. also exhibits no substantial change in comparisons with independent in-situ measurements \cite{liang_east_2022}. \\

This lack of change in the in situ comparisons may stem from two factors: \textbf{(a) Site insensitivity to changing emissions estimates from China.} In-situ observations used in this study include measurements from aircraft and surface towers, with details in SI Figure \ref{fig:SI-insitu_measurements}. Of the seven surface towers, only two are located in inland regions, whereas the remaining coastal towers are strongly influenced by pronounced shoreline-induced atmospheric dispersion. Of these inland towers, just one is situated in mainland China, operating as a regional background station (Mt Waliguan, 3,810 m above sea level) \cite{noaa_wlg_site}. Consequently, observations from these tower sites are likely to be  insensitive to regional emissions in China (i.e., Eastern \& Central China); \textbf{(b) Improved quality of emissions inventories.} Our analysis reveals a $\sim$7\% difference in magnitude between our posterior estimates and existing emissions inventories (see Section \ref{subsec1} in the main text). This modest adjustment to the overall emissions magnitude may not produce a noticeable change in the modeled mixing ratios at the in situ site, especially given that most of these sites are primarily designed to monitor background mixing ratios.

\newpage

\section{Supplementary Figures}

\begin{figure}[htb!]
    \centering
    \includegraphics[width=0.85\linewidth, trim=0 25 0 0, clip]{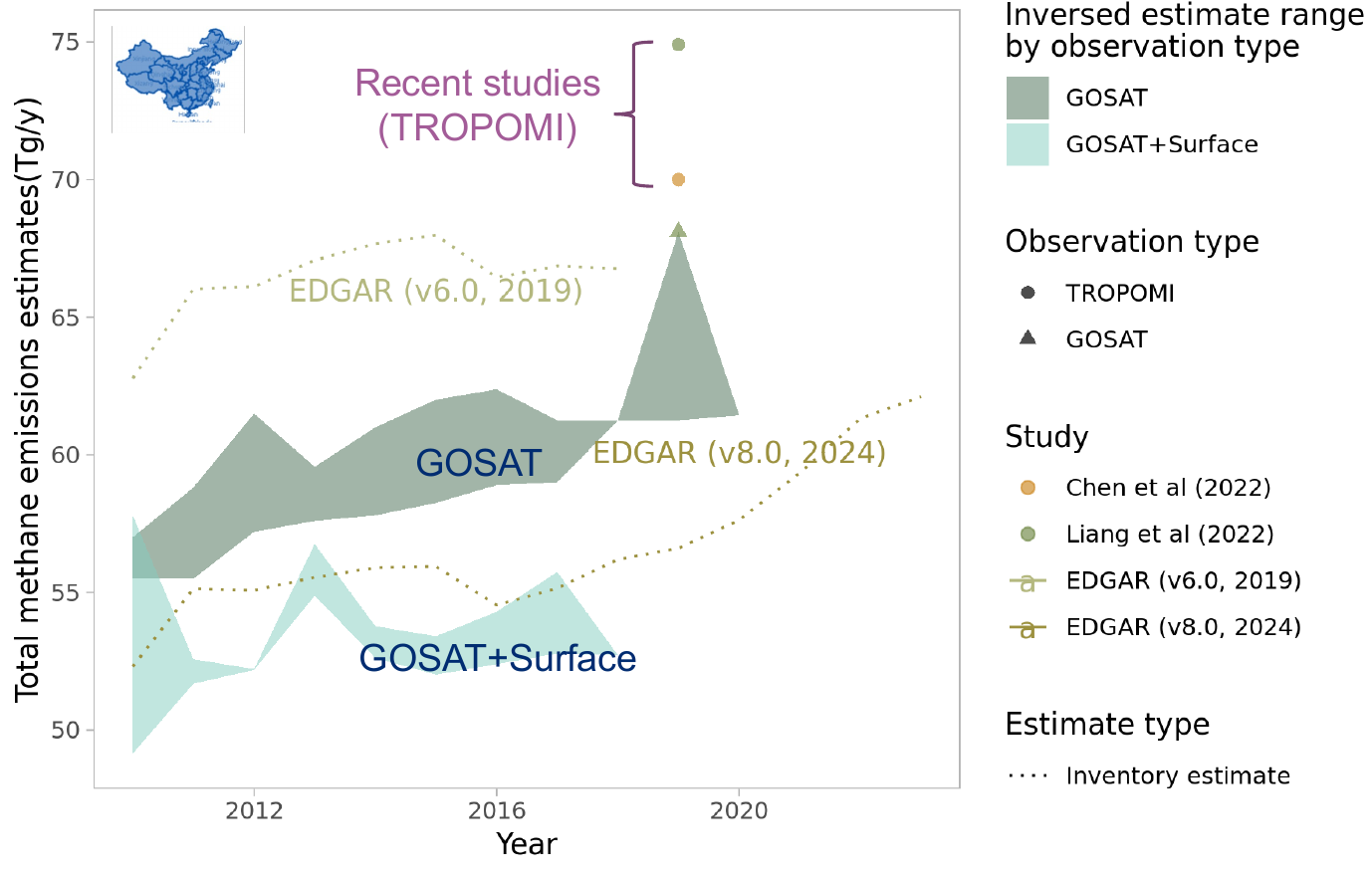}
    \caption{\textbf{Methane emissions estimates from China (2010--2019) show increased trends despite uncertainties:} Various regional inverse models show discrepant methane emissions estimates in 2010--2018, $\sim$10 Tg yr$^{-1}$ difference among studies with GOSAT (and surface) observations \cite{chen_methane_2022, liang_east_2022, zhang_observed_2022, zhao_slowdown_2024, miller_chinas_2019}. Most results from inverse models show lower \ce{CH_4} estimates from China than EDGAR v6.0 \cite{Crippa2021GHG}. In 2019, TROPOMI-based estimates are higher than GOSAT-based estimates. The newer EDGAR v8.0 suggests a climbing emissions trend from China after 2020 \cite{crippa_insights_2024}. The jury is still out on why studies yield somewhat different estimates for total emissions, but possible causes include (a) the specific research time window, (b) the selected sectoral inventories as prior information, (c) global/regional coverage and spatial resolution, (d) the chemical-transport model types/configurations, and (e) the specific satellite and/or surface measurements used (e.g., \cite{jacob2022review_quantify, behrendt2025USChinaUncertainty}). Comparing results from different models helps build confidence in understanding the overall emissions picture.}
    \label{fig:SI-uncertainty2010-2019}
\end{figure}


\begin{figure}[htb!]
    \centering
    \includegraphics[width=0.7\linewidth, trim=0 25 0 60, clip]{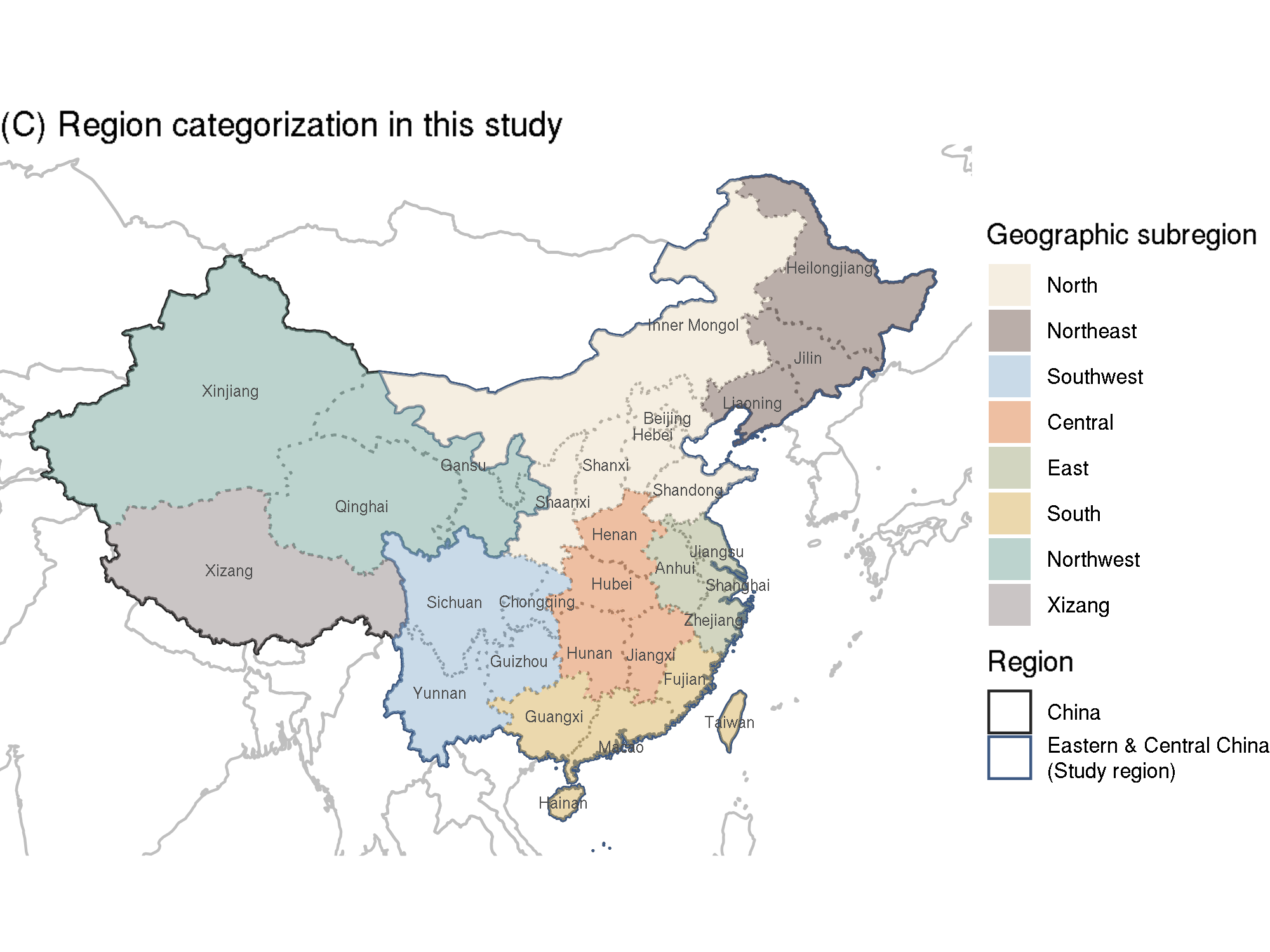}
    \caption{\textbf{Six sub-national regions within the study domain:} We analyze results in the Eastern \& Central China, which is comprised of the North, Northeast, Southwest, Central, East, South regions.}
    \label{fig:regioncat}
\end{figure}


\begin{figure}[htb!]
    \centering
    \includegraphics[width=1.0\linewidth]{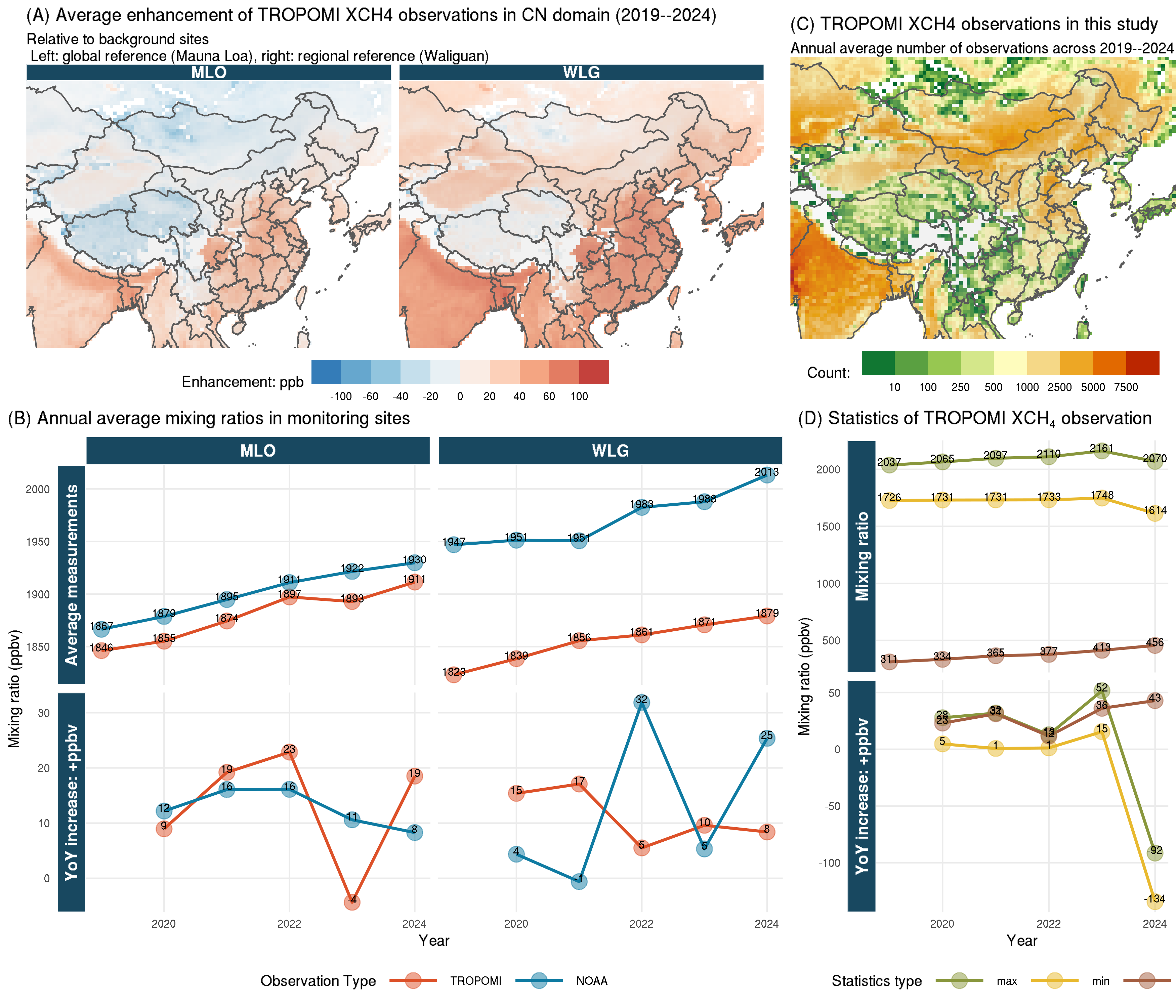}
    \caption{\textbf{TROPOMI \ce{XCH_4} observations steadily increase within the modeling domain after 2019:} (A) Average \ce{XCH_4} enhancement during 2019--2024, relative to global background (Mauna Loa - MLO) and regional background in East Asia (Waliguan - WLG); (B) Trend of annual mean \ce{XCH_4} and in-situ surface measurements for both MLO and WLG sites, along with associated year-over-year increase in the absolute value (with ppbv unit); (C) Spatial distribution of average annual occurrences of TROPOMI \ce{CH_4} observations during 2019--2024; (D) Trend of some statistics about monthly average \ce{CH_4} observations ($\mathrm{0.5^{\circ} \times 0.625^{\circ}}$) in each year: maximum, minimum, range, as well as the respective year-over-year (YoY) changes in ppbv.}
    \label{fig:SI-tropomi_observations}
\end{figure}

\begin{figure}[htb!]
    \centering
    \includegraphics[width=1.0\linewidth, trim=0 0 0 0, clip]{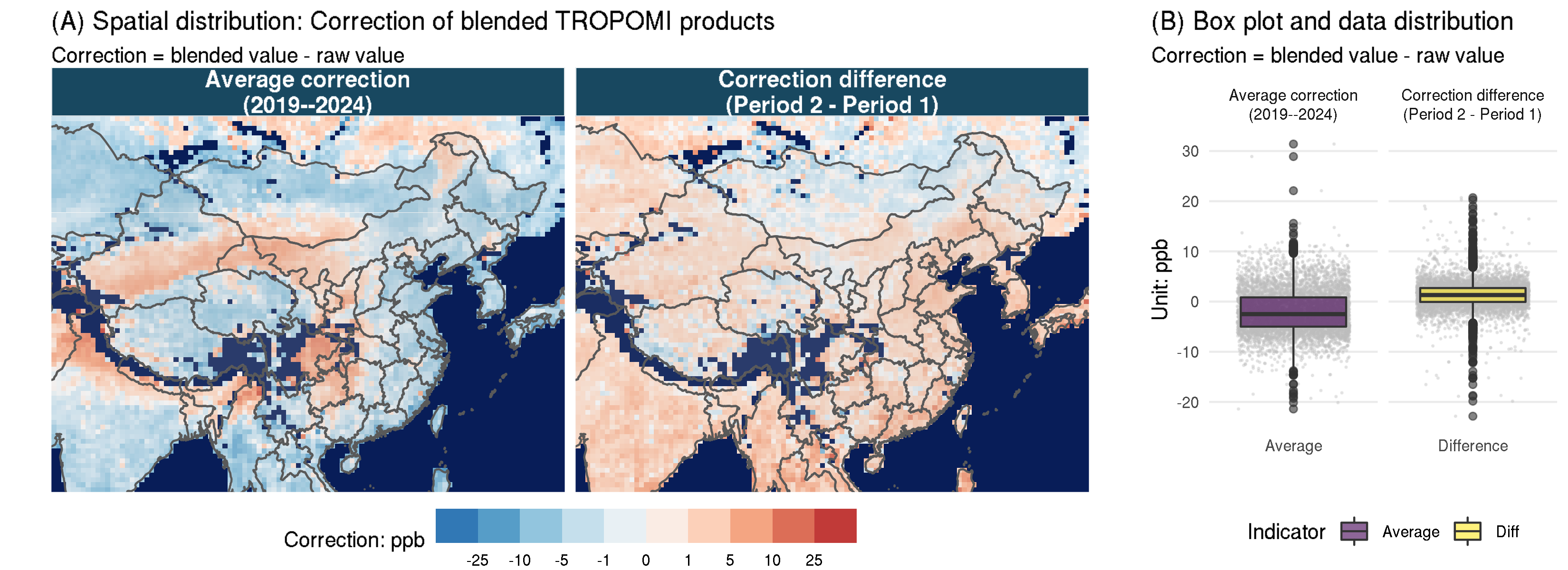}
    \caption{\textbf{Correction of blended TROPOMI+GOSAT satellite data product over China:} (A) Spatial distribution of mean correction from original TROPOMI products during 2019--2024 (left) and correction difference in Period 2 from Period 1 (right); Period 1 covers the years 2019--2021, while Period 2 covers the years 2022--2024; (B) Box plots and distributions of mean correction during 2019--2024 and the correction change in Period 2 from Period 1, where the box spans the 25$\mathrm{^{th}}$ to 75$\mathrm{^{th}}$ percentiles. }
    \label{fig:SI-correction_tropomi}
\end{figure}

\begin{figure}[htb!]
    \centering
    \includegraphics[width=\linewidth]{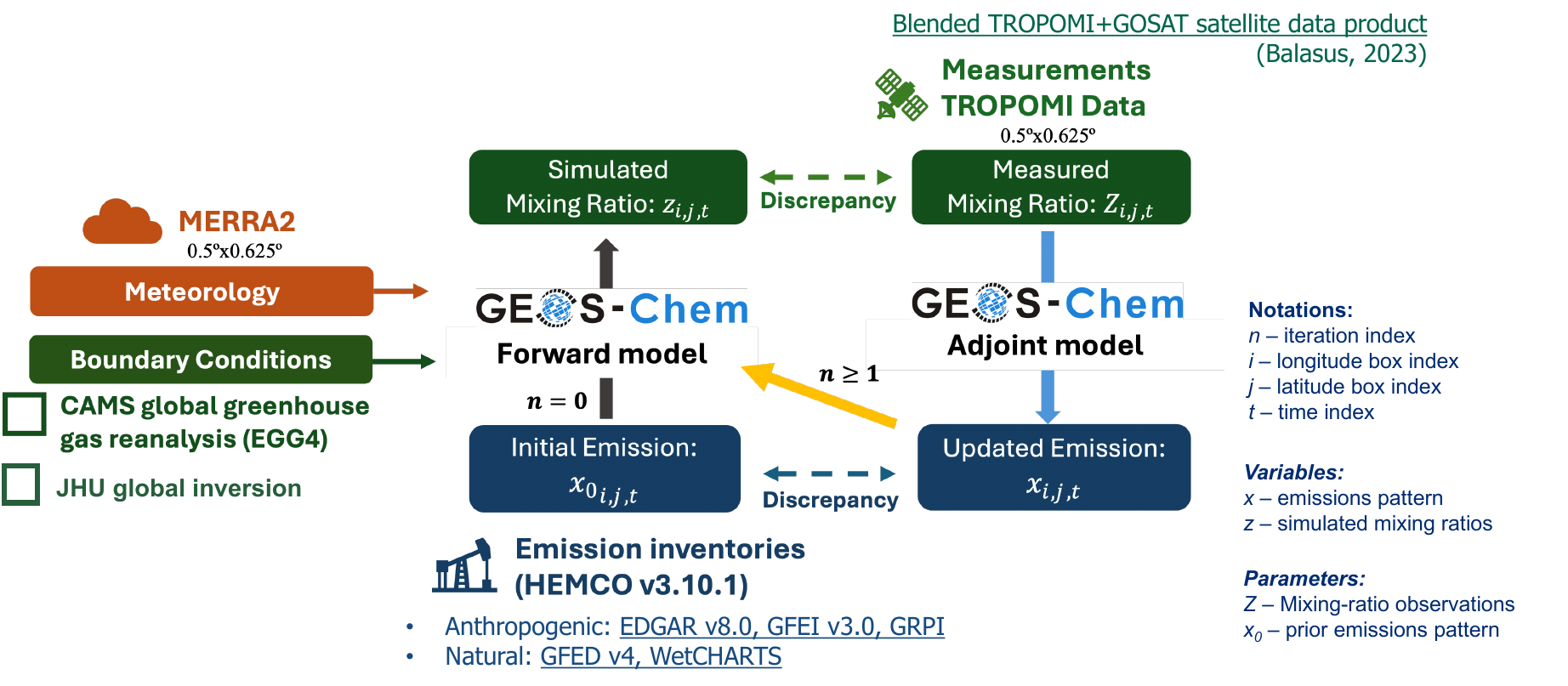}
    \caption{\textbf{Inverse model in this study optimizes emissions using a numerical approximation approach}}
    \label{fig:gim}
\end{figure}

\begin{figure}[htb!]
    \centering
    \includegraphics[width=1.0\linewidth, trim=0 0 0 20, clip]{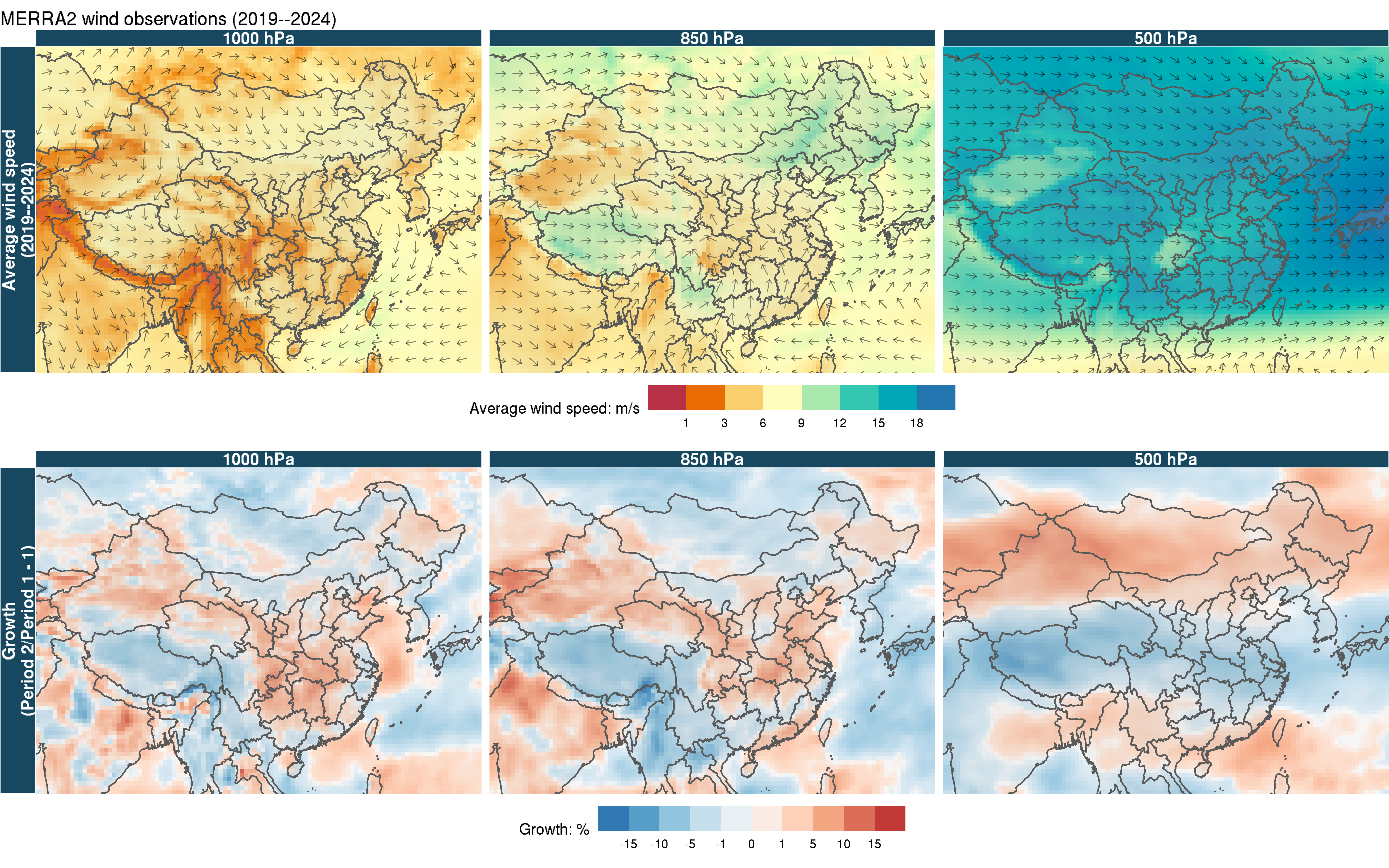}
    \caption{\textbf{MERRA2 wind observations over China across pressure layers:} The panels are wind maps across three pressure layers: 1000 hPa, 850 hPa, and 500 hPa (from left to right columns). Top row depicts the average wind speed across 2019--2024, with arrows representing wind direction. Bottom row shows the change of average wind speeds in Period 2 from Period 1, represented in percentage.}
    \label{fig:SI-wind_observations}
\end{figure}

\begin{figure}
    \centering
    \includegraphics[width=1\linewidth]{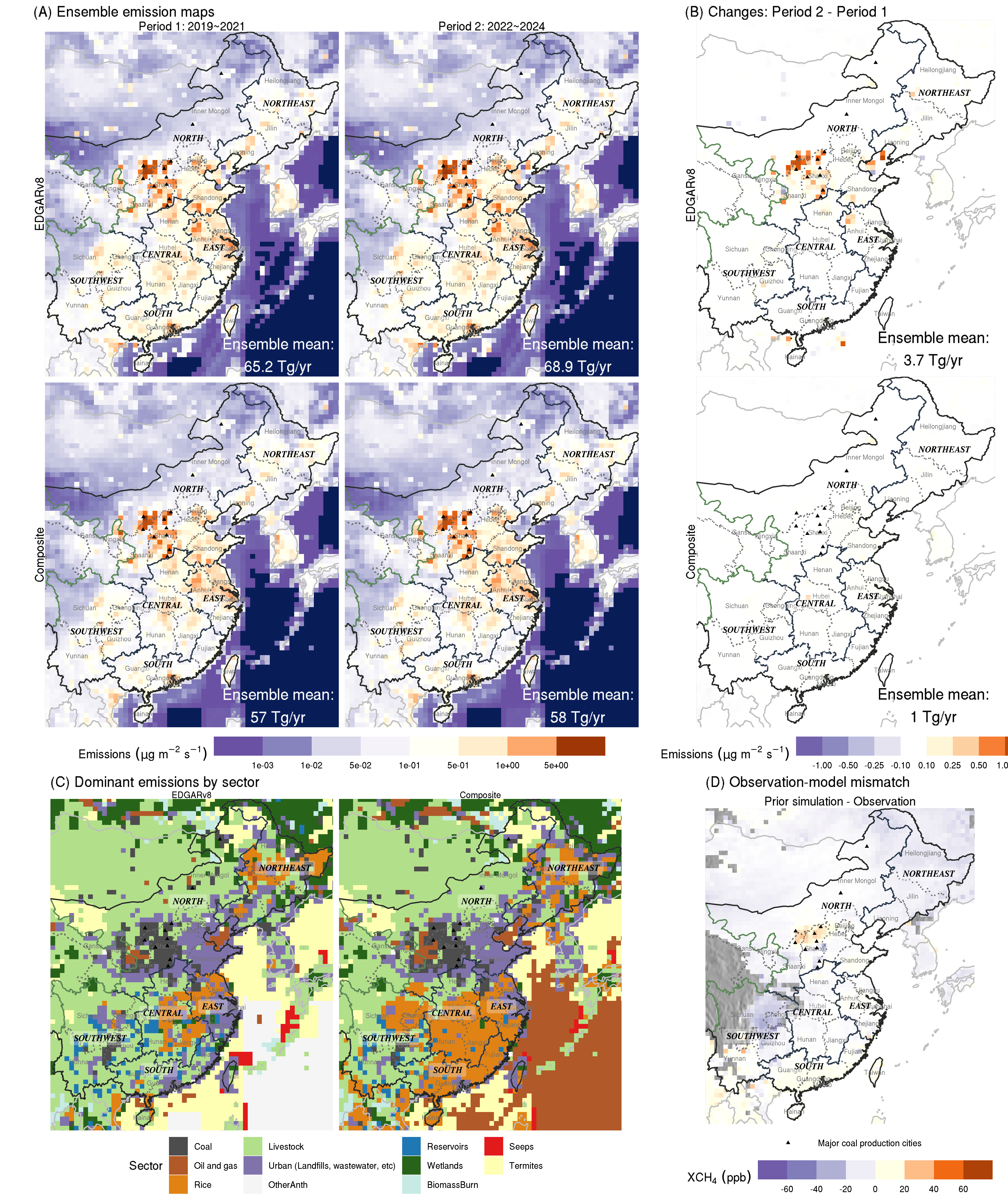}
    \caption{\textbf{Emission inventories, dominant sector based on emission inventories, and observation-model mismatch:} (A) Estimates of emission inventories in Period 1 and Period 2; (B) Variations of inventory estimate in Period 2 relative to Period 1; (C) Sectoral dominance based on the annual aggregation from emission inventories; (D) Mismatch between prior simulation with emission inventories and TROPOMI \ce{CH_4} observations in ppbv.}
    \label{fig:SI-emissions_inventories}
\end{figure} 

\begin{figure}[htb!]
    \centering
    \includegraphics[width=1.0\linewidth, trim=0 0 0 20, clip]{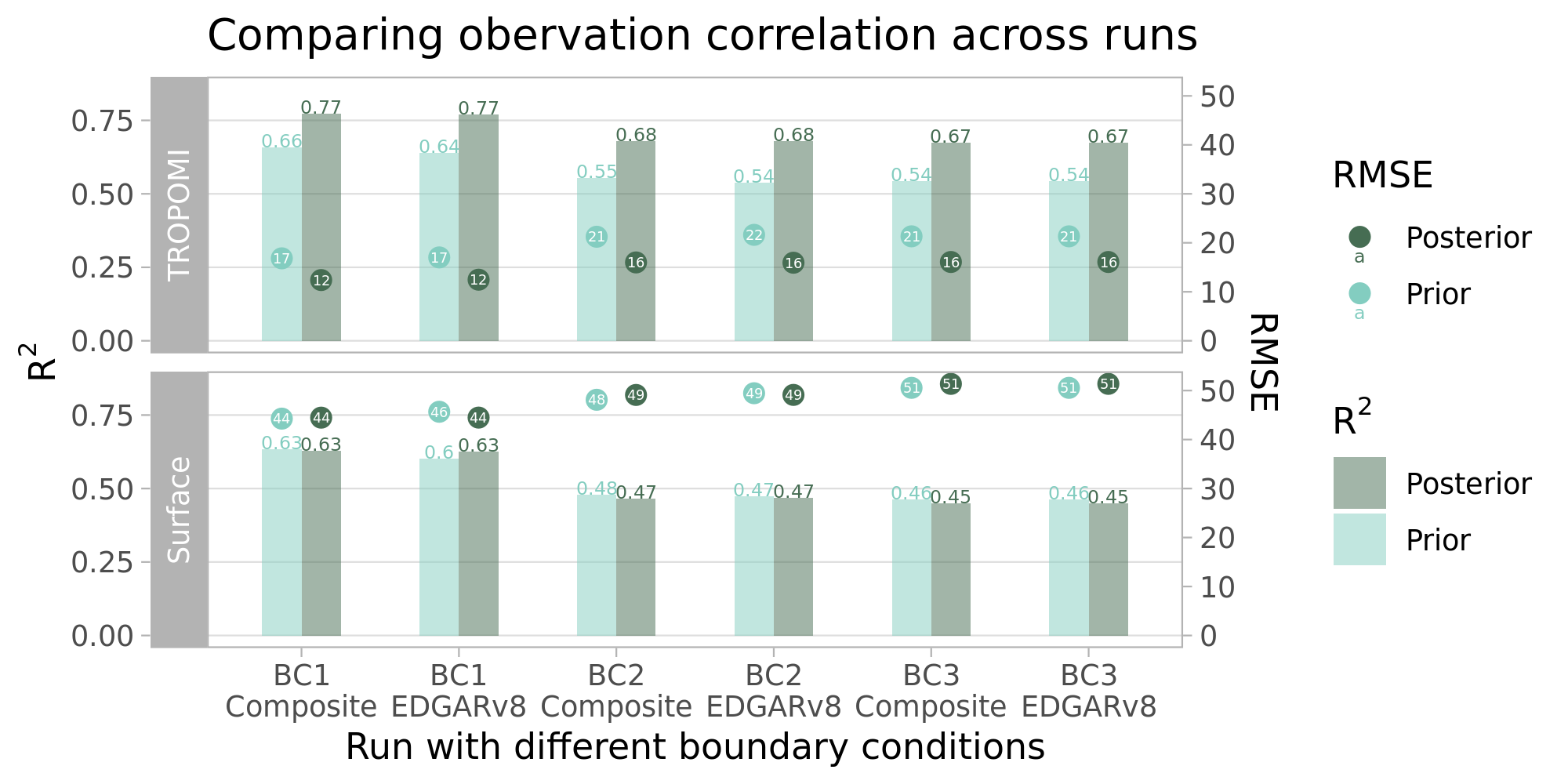}
    \caption{\textbf{Posterior simulations align better with TROPOMI \ce{XCH_4} observations:} The plot shows the comparison between prior and posterior model simulation against measurements, including TROPOMI (top) and in-situ observations (bottom). The in-situ network includes aircraft observations (Cape Ochi-ishi, CON; In-service Aircraft for a Global Observing System, IAGOS) and surface tower observations (Mount Waliguan, MLG; Dongsha Island, DSI; Lulin, LLN; Ulaan Uul, UUM; Yonagunijima, YON; Anmyeon-do, AMY; Tae-ahn Peninsula, TAP).}
    \label{fig:SI-insitu_measurements}
\end{figure}

\begin{figure}[htb!]
    \centering
    \includegraphics[width=\linewidth]{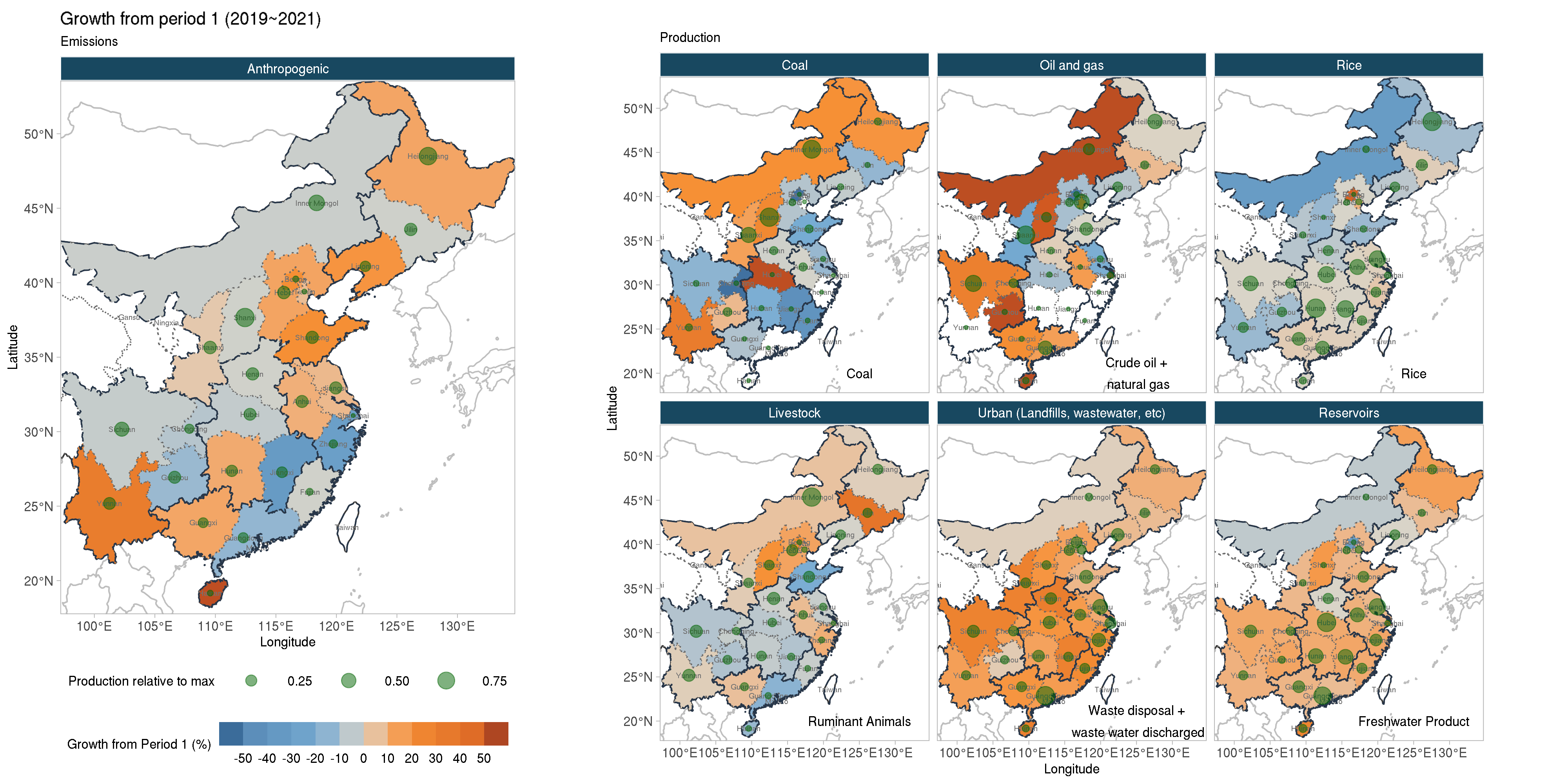}
    \caption{\textbf{Provincial \ce{CH4} emissions variation and production activity changes in Eastern and Central China:} The plots depict the percentage growth of Period 2 (2022--2024) from Period 1 (2019--2021) in emissions and production of primary activities in each sector. Colors show the growth of average production in Period 2, compared to Period 1. The size of each green bubble indicates the production level of a province, relative to the province with the highest production.}
    \label{fig:emissions_activity_trend}
\end{figure}

\begin{figure}[htb!]
    \centering
    \includegraphics[width=1\linewidth, trim=0 0 0 0, clip]{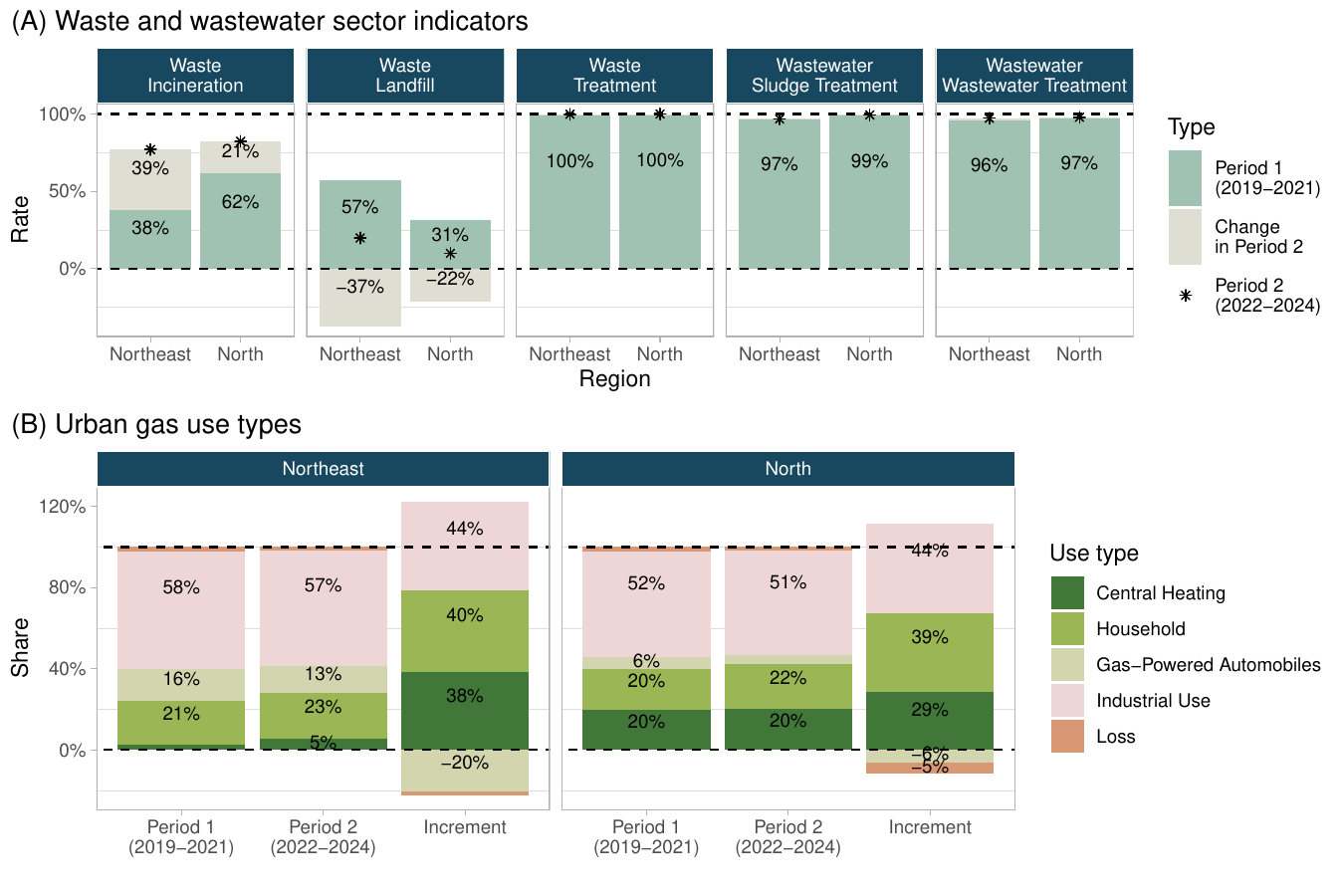}
    \caption{\textbf{Comparison of urban activities in waste, and wastewater and urban gas use across North and Northeast regions:} (A) Changing treatment rate of waste and wastewater sector in North and Northeast. As listed from left to right, in the waste sector, incineration and landfill rates are the respective volume ratios to the total treated waste, while the waste treatment rate is the volume ratio of treated waste to total waste. In the wastewater sector, the treatment rates are the treatment volumes ratios to sludge production or total wastewater discharge. Both the Northern and Northeastern regions exhibit very high rates of solid waste and wastewater treatment and have progressively transitioned from landfilling to incineration as the primary method for urban waste disposal; (B) Urban gas use mix in North and Northeast region. The combined share of urban gas use for central heating and household slightly increase in Period 2 from Period 1. The incremental gas consumption in Period 2 is mainly driven by central heating and household use, accounting for much higher share than in the existing consumption mix. }
    \label{fig:SI_urban}
\end{figure}


\begin{figure}[htb!]
    \centering
    \includegraphics[width=1\linewidth, trim=0 50 0 75, clip] {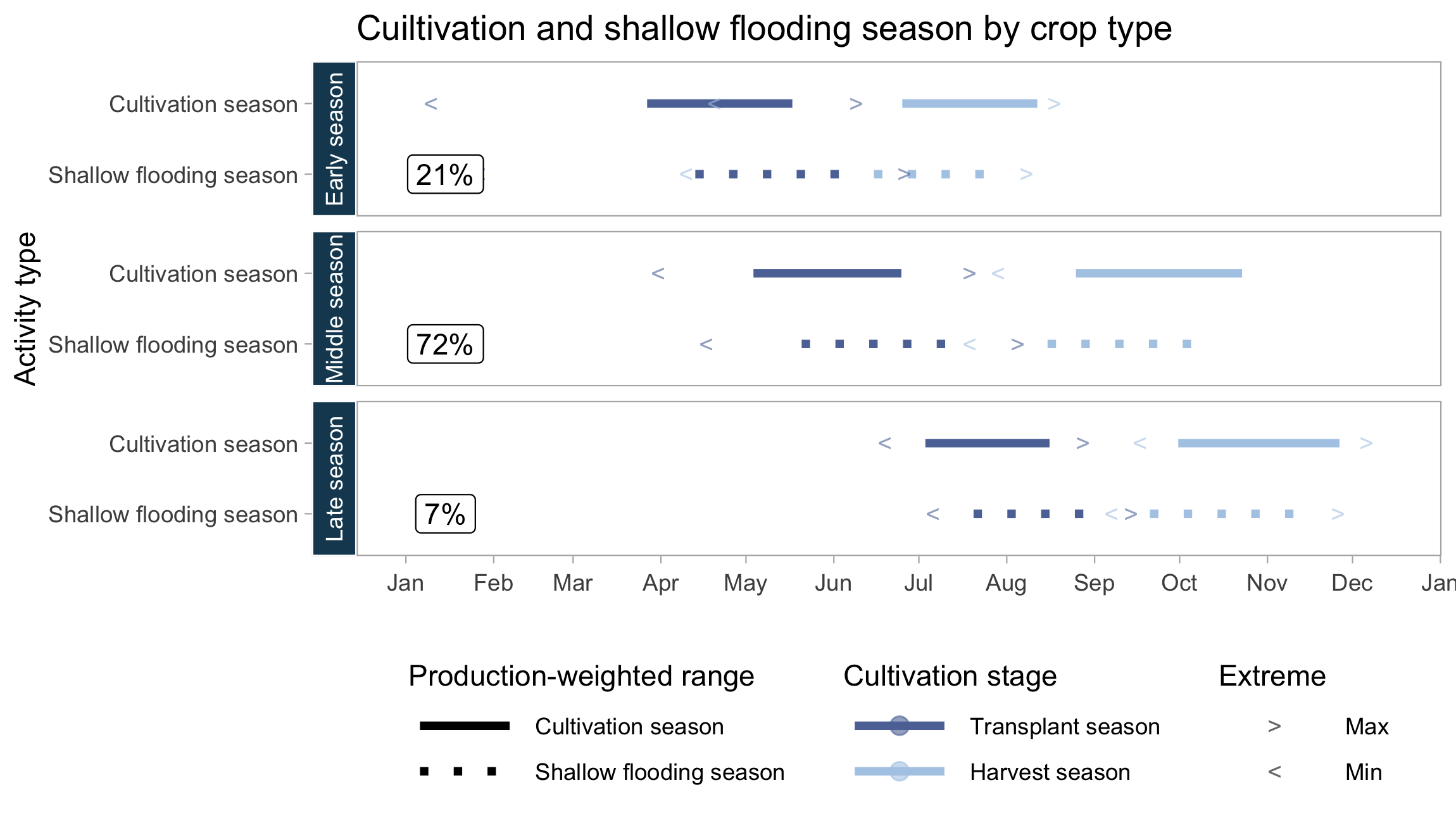}
    \caption{\textbf{Weighted seasonality in cultivation and shallow flooding by crop type: } Labels are estimated average production shares of early, middle, and late-season crops during 2019--2024. We calculate the average date for each crop type, based on the provincial calendar proposed by Li et al and weighted by provincial production \cite{li2024chinaricecalendar}. The provincial rice production data (2019--2024) is from National Bureau of Statistics (NBS). NBS reports  yields for early, middle, and double-season crops. We assume that the yields of double-season crops halve in early and late seasons. We also estimate that flooding begins approximately 17 days after the transplant date and ends roughly 10 days before the harvest date.} 
    \label{fig:SI_rice}
\end{figure}

\clearpage

\section{Supplementary Tables}
\begin{table}[ht!]
    \centering
    \caption{Ensemble sectoral emissions in Eastern \& Central China \\ ($\mathrm{Tg\ y^{-1}}$, 2019--2024)}
    \label{tab:placeholder_label}
    \begin{tabular}{llccccc}
        \toprule
        Sector & 
        \makecell[c]{Sector\\type} & 
        \makecell[c]{This study\\(Ensemble mean)} & 
        \makecell[c]{Inventory set\\ (EDGARv8) \\ Prior A}  &
        \makecell[c]{Inventory set\\ (Composite) \\ Prior B}  &
        \makecell[c]{Referenced\\global study} \\
        \midrule
        Anthropogenic & Level1 & 47.4 (86.9\%) & 63.7 (93.2\%) & 54.1 (92\%) & 40.2 (91.9\%) \\
        \hline
        Coal          & Level2 & 10.7 (22.5\%) & 22.6 (35.5\%) & 19 (35.1\%) & 13.2 (32.8\%) \\
        Oil and gas   & Level2 & 1.4 (2.9\%)   & 1.3 (2.1\%)   & 0.9 (1.7\%) & 0.8 (2.1\%)  \\
        Landfills and was & Level2 & 12.3 (25.9\%) & 14.9 (23.4\%) & 15 (27.7\%) & 9.1 (22.7\%)  \\
        Livestock     & Level2 & 10.6 (22.4\%) & 7.7 (12.1\%)  & 7.8 (14.4\%) & 4.9 (12.2\%) \\
        Rice          & Level2 & 9 (19\%)      & 13.7 (21.5\%) & 7.9 (14.6\%) & 9.1 (22.6\%) \\
        OtherAnth     & Level2 & 3.5 (7.4\%)   & 3.5 (5.4\%)   & 3.5 (6.4\%) & 3 (7.6\%)    \\
        \bottomrule
    \end{tabular}
    \footnotetext{Note: For level-1 sector types, the percentage for each sector indicates its portion of total emissions, while for level-2 sector types, it indicates its portion of anthropogenic emissions. The referenced global study is Pendergrass et al. (2025). }
\end{table}

\begin{table}[h]
    \centering
    \caption{Production of top coal-mining cities in our study region}
    \label{tab:coal_cities}
    \begin{tabular}{llcccc}
        \hline
             &    & \multicolumn{3}{c}{Average raw coal production (mt)} & Growth rate \\
        City &  Province & 2019--2023 & \makecell{Period 1\\2019--2021} & 
             \makecell{Period 2\\2022--2023} & Period 2 from Period 1 \\
        \hline
        Ordos      & Inner Mongolia& 756.8 & 693.3 & 820.4 & 18\%  \\
        Yulin      & Shaanxi      & 552.9 & 511.0 & 594.9 & 16\%  \\
        Shouzhou   & Shanxi       & 206.7 & 188.4 & 225.0 & 19\%  \\
        Changzhi   & Shanxi       & 156.0 & 141.8 & 170.3 & 20\%  \\
        Lyuliang   & Shanxi       & 145.9 & 135.2 & 156.7 & 16\%  \\
        Datong     & Shanxi       & 149.4 & 134.9 & 164.0 & 22\%  \\
        Jincheng   & Shanxi       & 130.8 & 118.5 & 143.1 & 21\%  \\
        Xilinguole & Inner Mongolia& 125.2 & 114.0 & 136.4 & 20\%  \\
        Jinzhong   & Shanxi       & 107.9 & 93.5  & 122.3 & 31\%  \\
        Hulunbeier & Inner Mongolia& 99.3  & 92.1  & 106.5 & 16\%  \\
        Xinzhou    & Shanxi       & 108.6 & 84.8  & 132.3 & 56\%  \\
        \hline
    \end{tabular}
    \footnotetext{Data source: Annual city-level raw coal production data is from China Coal Industry Yearbook (2023) \cite{ChinaCoalYB-2023} and from the statistical yearbooks of Shanxi (2019--2023) \cite{ShanxiYB-2023}, Inner Mongolia (2019--2023) \cite{InnerMongoliaYB-2023} and Yulin City (2024) \cite{YulinYB-2024}.}
\end{table}

\begin{table}[h]
    \centering
    \caption{Agricultural production within our study region (Millions, 2019--2024)}
    \label{tab:ag_products}
    \begin{tabular}{llllllllll}
        \hline
Unit & Indicator & Type & 2019 & 2020 & 2021 & 2022 & 2023 & 2024 & Growth \\
\midrule
Tons & Production & Milk & 27 & 29 & 30 & 32 & 33 & 32 & 19\% \\
Tons & Production & Beef & 7 & 7 & 7 & 7 & 8 & 8 & 17\% \\
\hline
Heads & Yr-end stock & Ruminants & 261.7 & 269.4 & 276.8 & 282.7 & 286.1 & 266.1 & 2\% \\
Heads & Yr-end stock & Ruminants.ceq & 72.5 & 76.8 & 79.2 & 82.9 & 89.0 & 83.0 & 15\% \\
Heads & Yr-end stock & Cattle & 69.4 & 71.0 & 71.8 & 74.5 & 75.4 & 70.2 & 1\% \\
Heads & Yr-end stock & \% Cattle & 23\% & 23\% & 24\% & 24\% & 26\% & 26\% & \\
Heads & Yr-end stock & \% Cattle.ceq & 71\% & 72\% & 72\% & 73\% & 75\% & 76\% & \\
\hline
Heads & Annual sales & Ruminants & 284.0 & 286.5 & 295.4 & 301.5 & 298.4 & 280.9 & -1\% \\
Heads & Annual sales & Ruminants.ceq & 61.6 & 61.7 & 63.4 & 65.0 & 65.5 & 63.8 & 4\% \\
Heads & Annual sales & Cattle & 36.9 & 36.7 & 37.6 & 38.7 & 39.6 & 39.7 & 8\% \\
Heads & Annual sales & \% Cattle & 13\% & 13\% & 13\% & 13\% & 13\% & 14\% & \\
Heads & Annual sales & \% Cattle.ceq & 60\% & 60\% & 59\% & 60\% & 60\% & 62\% & \\
\hline
Heads & Avg. stock & Ruminants & 411.2 & 420.4 & 432.4 & 441.8 & 443.6 & 414.0 & 1\% \\
Heads & Avg. stock & Ruminants.ceq & 103.3 & 107.6 & 110.9 & 115.4 & 121.8 & 115.0 & 11\% \\
Heads & Avg. stock & Cattle & 69.4 & 71.0 & 71.8 & 74.5 & 75.4 & 70.2 & 1\% \\
Heads & Avg. stock & \% Cattle & 19\% & 19\% & 19\% & 20\% & 21\% & 22\% & \\
Heads & Avg. stock & \% Cattle.ceq & 68\% & 69\% & 69\% & 69\% & 71\% & 72\% & \\
\bottomrule
    \end{tabular}
    \footnotetext[1]{Milk and beef product data is aggregated based on the annual provincial data from National Bureau of Statistics of China \cite{ProvinceStats-2019-2024}.}
    \footnotetext[2]{Livestock-related data comes from the China Rural Year Book (2020--2025) \cite{china_rural_yb_2020_2025}, which reports stocks and sales of cattle and sheep in the previous year. In the table, the abbreviations used are ``yr-end stock'' for year-end stock and ``sales'' for annual sales. Annual average stock (avg. stock) is computed using formula: $\mathrm{Stock_{Average} = Stock_{Year-end} +  0.5 \times Sales_{Annual}}$, where we assume sales are evenly distributed in a year.} 
    \footnotetext[3]{We also calculate the ruminant animals in ``cattle equivalent''(ceq) in terms of methane emissions, where emissions of sheep is equivalent to 0.1 of the emissions of cattle (1 sheep = 0.1 cattle equivalent) \cite{broucek2014livestockmethane}. In this table, ruminant animals are limited to cattle and sheep for comparison, as provincial annual camel sales data is not available.}
    \footnotetext[4]{Share of cattle is the ratio of cattle to total ruminants, measured both in headcount and in cattle equivalents.}
\end{table}

\end{appendices}

\newpage
\putbib[01-sn-bibliography]

\end{bibunit}

\clearpage

\end{document}